\documentclass[12pt]{iopart}

\expandafter\let\csname equation*\endcsname\undefined
\expandafter\let\csname endequation*\endcsname\undefined
\usepackage{amsmath}
\usepackage{microtype}
\usepackage{graphicx}
\usepackage{subfigure}
\usepackage{booktabs} 
\usepackage{braket}

\usepackage{hyperref}

\usepackage{amssymb}
\usepackage{mathtools}
\usepackage{amsthm}
\usepackage{color}

\usepackage[capitalize,noabbrev]{cleveref}

\theoremstyle{plain}

\theoremstyle{definition}

\theoremstyle{remark}

\begin{document}
\title[Machine-learning emergent spacetime from linear response in future tabletop quantum...]{Machine-learning emergent spacetime from linear response in future tabletop quantum gravity experiments}


\author{Koji Hashimoto$^1$, Koshiro Matsuo$^2$,
Masaki Murata$^2$, Gakuto Ogiwara$^2$ and Daichi Takeda$^1$}
\address{$^1$ Department of Physics, Kyoto University, Kyoto 606-8502, Japan}
\address{$^2$ Department of Information Systems, Saitama Institute of Technology, Saitama 369-0293, Japan}
\ead{koji@scphys.kyoto-u.ac.jp, f3011hxs@sit.ac.jp, m.murata@sit.ac.jp, f3002keg@sit.ac.jp, takedai@gauge.scphys.kyoto-u.ac.jp}

\vspace{10pt}

\def\time{\Xi}

\begin{abstract}
We introduce a novel interpretable Neural Network (NN) model designed to perform precision bulk reconstruction under the AdS/CFT correspondence.
According to the correspondence, a specific condensed matter system on a ring is holographically equivalent to a gravitational system on a bulk disk, through which tabletop quantum gravity experiments may be possible as reported in \cite{hashimoto2022spacetimeemergent}. The purpose of this paper is to reconstruct a higher-dimensional gravity metric from the condensed matter system data via machine learning using the NN.
Our machine reads spatially and temporarily inhomogeneous linear response data of the condensed matter system, and incorporates a novel layer that implements the Runge-Kutta method to achieve better numerical control.
We confirm that our machine can let a higher-dimensional gravity metric be automatically emergent as its interpretable weights, using a linear response of the condensed matter system as data, through supervised machine learning. The developed method could serve as a foundation for generic bulk reconstruction, {\it i.e.}, a practical solution to the AdS/CFT correspondence, and would be implemented in future tabletop quantum gravity experiments.
\end{abstract}

\section{Introduction}

\label{sec:Intro}
In this paper, we provide a concrete physical neural network model addressing one of the long-standing issues in quantum gravity: the bulk reconstruction in holography, or 
the AdS/CFT correspondence \cite{Maldacena_1999}. 
The holographic correspondence posits that two quantum theories --- non-gravitating quantum field theory (named ``CFT side") and gravity theory (named ``AdS side") --- are equivalent.
The CFT side is called boundary and the AdS side is called bulk, because the gravity theory is defined on a ($d+1$)-dimensional spacetime manifold, whose boundary is isomorphic to the $d$-dimensional manifold where the CFT side is defined.
Although no formal proof for the correspondence has yet been provided so far, the conjecture is supported by numerous practical examples.

A primary challenge confronting this conjecture is the lack of a constructive approach to deduce the background geometry in the gravity theory from some data of the boundary non-gravitating quantum field theory.
This is called bulk reconstruction and it will help us understand other essential questions, such as the identification of the gravity theory dual to a given boundary theory, or to what extent of theories the holographic duality is valid.
The algorithm for the bulk reconstruction could be implemented for future tabletop material experiments for emergent spacetime, as a central engine.

In this paper, we use the linear response as the boundary data to reconstruct the bulk geometry.
This is because the linear response theory is practical if one has future experiments in mind (such as the one provided in \cite{hashimoto2022spacetimeemergent}), and furthermore, it is also theoretically reasonable in the sense that we do not need to assume much about the bulk theory such as the gravity action, which is generally not necessarily the same as the Einstein's general relativity.
We consider acting source $J$ on a scalar operator $O$ of the boundary theory, and measure the expectation value $\braket{O}_J$ to the linear order in $J$.
The linear response data of the boundary theory can be calculated from the gravity theory if the duality holds, and according to the holographic dictionary, it is equivalent to simply solving the equation of motion (EOM) of the bulk scalar field on a fixed background geometry, if the bulk metric is known.
Usually in the research of the holographic correspondence, a bulk geometry background is predetermined by inference and we solve the EOM to predict the response $\braket{O}_J$ from the input source $J$.
Conversely, in this paper, we begin with a given set of spacetime dependent $J$ and $\braket{O}_J$, and try to find the bulk background geometry (metric) that appears as the coefficients of the bulk scalar EOM.
In short, the measurement on a material provides the linear response data, and our goal is to spot the unknown coefficients of the bulk scalar EOM so that the bulk solution reproduces the boundary material data.

A simple and realistic setup for addressing this problem in holography is the case where the bulk is 3-dimensional and the boundary is 2-dimensional. 
Since we are interested in the linear response, 
the bulk degree of freedom is just fine with a single
free scalar field propagating over a fixed background spacetime.\footnote{In \cite{Fan:2023bsz}, the authors proposed to reconstruct the two components of the metric using two scalar fields with different masses.
While they picked up the $s$-wave components, we challenge here, in the light of the experimental availability, to build the metric with only a single field, but with the help of higher wavenumbers.}
We follow the scenario outlined in \cite{hashimoto2022spacetimeemergent}, where the space of the boundary theory is a one-dimensional circle (ring), which is a central target of the tabletop quantum gravity experiment in a laboratory with condensed matter.
The background in the bulk theory can be assumed static and rotationally symmetric, because it is reasonable to suppose that the condensed matter is prepared in a symmetric equilibrium state.
As proposed in \cite{hashimoto2018deep}, we build a sparse neural network that models the bulk differential equation, ensuring interpretability by making the weights correspond to the EOM coefficients, from which we are able to recover the bulk background metric.\footnote{
Gravitational spacetime itself is also represented by a deep neural network in \cite{you2018machine}. See also subsequent advancements in \cite{hashimoto2018deep2,hashimoto2019ads,vasseur2019entanglement,tan2019deep,akutagawa2020deep,hu2020machine,yan2020deep,hashimoto2021neural,lam2021machine,song2021ads,yaraie2021physics,li2023learning,ahn2024deep,mansouri2024holographic}. In particular, \cite{ahn2024deep} demonstrates machine learning of an emergent spacetime from the conductivity data of a real material. For the recent use of machine learning architecture for pinning down a holographic metric without resorting to identifying the neural network as a spacetime, see for example \cite{chen2024machine, luo2024neural, chen2024flavor, ahn2024holographic,gu2024neural,bea2024gravitational}. 
}

This paper is organized as follows.
In the next section, we review two basic concepts essential to this work: the AdS/CFT correspondence and spacetime-emergent material.
In Section 3, we describe our methodology, including the structure of our neural network and the process for creating the dataset.
The numerical results of the machine learning are presented in Section 4, followed by a discussion in Section 5.

\section{Brief review of holography and its experiment}

\subsection{The AdS/CFT Correspondence}

\label{sec:AdS/CFT}

The AdS/CFT correspondence \cite{Maldacena_1999} is a conjecture which states that a quantum gravity theory with a negative cosmological constant in $d+1$-dimensional asymptotically AdS (anti-de Sitter) spacetime is equivalent to a non-gravitating quantum field theory in $d$-dimensional spacetime. The former is referred to as the bulk theory, while the latter as the boundary theory. The AdS is the maximally symmetric spacetime of constant negative curvature. Although the conjecture has not been theoretically proven yet, various pieces of evidence have been found, and its application is now widespread, including in fields such as condensed matter physics and quantum computation.

With a given bulk gravity theory, in its classical limit we can compute the correlators on the dual boundary theory straightforwardly. All that is necessary is to solve equations of motion (EOMs) under specific boundary conditions to calculate the bulk on-shell action. This on-shell action, acording to the conjecture, corresponds to the generating functional for the boundary theory correlators. On the other hand, through this correspondence, the inverse problem --- constructing bulk fields and the bulk geometry from the boundary theory data --- is generally challenging, because it requires identifying (reconstructing) the EOM that is consistent with a given set of the boundary theory correlators. The difficulty of the inverse problem is in the difference of the spatial dimensions in holography. This issue is known as ``bulk reconstruction" and is a well-posed problem situated within the realm of physics working towards quantum gravity.

\subsection{Spacetime-emergent Material (SEM)}
\label{sec:SEM}

Following \cite{hashimoto2022spacetimeemergent} we call as SEM, a material whose properties can be equivalently described by a higher-dimensional quantum gravity theory under the AdS/CFT conjecture.
Below in this paper, we focus especially on a ring-shaped material, due to its ease in being realized in possible experiments.

According to the AdS/CFT correspondence, when the bulk is a black hole spacetime, the corresponding boundary theory is known to be a finite temperature QFT \cite{Witten:1998zw}. Given that generic condensed matter systems near a quantum critical point (QCP) are governed by a thermal CFT, it is hypothesized that a material allowing for a higher-dimensional gravity description can exist; this was named a SEM in \cite{hashimoto2022spacetimeemergent}. The simplest realization of the AdS/CFT in our world would be a ring-shaped thermal material near a QCP, with spacetime topology $\mathbb{S}^1 \times \mathbb{R}$. In this scenario, the dual emergent three-dimensional gravity theory is defined in the spatial region enclosed by the ring --- a disk.

Experimental verification of SEMs will aid in understanding the conjecture and unraveling the mysteries of quantum gravity. To achieve this, we need first to theoretically reconstruct the bulk from limited experimental data of the candidate material and verify whether the reconstructed bulk can predict other phenomena and remain consistent with subsequent experiments. Consequently, the initial step is to establish a universal theoretical method for determining the bulk metric from available experimental data. Our novel neural network is designed for this particular purpose.

\section{Method of machine learning}
\label{sec:method}

In this study, we consider a theoretical setup for a material experiment where a small external source is applied to some local operator in the theory and its linear response function is measured. Thus, the boundary data available to us are the values of the source and the response, which, according to the AdS/CFT dictionary, are related to the asymptotic behavior of a scalar field in a gravitational curved spacetime.\footnote{
We here consider that the boundary local operator is a scalar for simplicity.
When it is not, we must change its bulk counterpart to a field in a different representation of the Lorentz group.
}
To reconstruct the bulk metric, we use a neural network (NN) based approach equivalent to the Klein-Gordon equation of the scalar field on the unknown metric.

Let $\Phi(t,\theta,\xi)$ be the scalar field in the 3-dimensional bulk, where $t$ is the time coordinate, $\theta$ is the coordinate along $\mathbb{S}^1$, and $\xi \in [0,1]$ is the radial coordinate (the extra dimension unique to the bulk). 
It is always possible to choose $\xi$ such that $\xi = 1$ corresponds to the ring on which the dual CFT lives. 
We Fourier-expand the scalar field as $\Phi(t,\theta,\xi) = \sum_n \Phi_n(\xi) e^{-i\omega t + i k_n \theta}$, where $k_n = \frac{2\pi n}{a}$ with $a$ being the ring circumference, and $\omega$ is the frequency of the external source, which are
controllable in the experiment. Since our interest is in the forced oscillation part that remains after enough time has passed, only the component with frequency $\omega$ is taken into consideration. We introduce the conjugate of the scalar field $\Pi_n(\xi) = \Phi_n'(\xi)$, with which the Klein-Gordon equation is reduced to first-order equations. This is essential for the equation of motion to be rewritten as a NN architecture.

In the phase of SEM verification, it is reasonable to use a material that is static, rotationally symmetric, and in equilibrium. 
Assuming this situation, we can suppose that the bulk metric is of the following form:
\begin{align}{\label{eq:metric}}
ds^2 = g_{tt}(\xi)dt^2 + g_{\xi\xi}(\xi)d\xi^2 + g_{\theta\theta}(\xi)d\theta^2.
\end{align}
The Klein-Gordon equation with this metric is:
\begin{align}
0 = -\omega^2\time(\xi)\Phi_n(\xi) &+ \partial_\xi\left( \ln \sqrt{-g} g^{\xi\xi} \right)\Pi_n(\xi)+ \Pi'_n(\xi) - k_n^2 \Theta(\xi)\Phi_n(\xi).\label{eq: EOM before NN}
\end{align}
Here, we have introduced the product combinations of the metric:
\begin{align}
\time(\xi) = g_{\xi\xi} g^{tt}, \quad
\Theta(\xi) = g_{\xi\xi} g^{\theta\theta}.
\end{align}
By using the residual redundancy of the diffeomorphism, we gauge-fix the metric as:
\begin{align}
\sqrt{-g}g^{\xi\xi} = C\xi^{-1},\label{eq: gauge condition}
\end{align}
where any constant is allowed for $C$ and we will choose a value for it later. The validity of this gauge-fixing is explained in \ref{app:gauge-fix}, where we will also see how to recover the metric under this gauge condition. Thus, all we have to do to reconstruct the metric is to find $(\time(\xi), \Theta(\xi))$ from the data.

\subsection{Numerical Accuracy and discretization scheme}
\label{sec:Numerical analysis without NN}
In the original research \cite{hashimoto2018deep}, the EOM for the bulk scalar field is discretized by using the Euler method and is interpreted and identified as a NN. 
In that work, the input to the NN is the value of the scalar field and its conjugate at a specific point within the bulk. 
In contrast, our approach focuses on the data of the material side, which corresponds to the asymptotic behavior of the scalar field near the AdS boundary, based on the AdS/CFT dictionary.
The asymptotic value of the field and its conjugate have the equivalent information to the boundary dataset of $J$ and $\braket{O}_J$ (or, their Fourier modes more strictly).
In this case, the adequacy of the Euler method in providing sufficient accuracy is not clear, since the value of the bulk field grows up as it approaches to the boundary.

In this section, we demonstrate that the Euler method introduces significant errors, while the Runge-Kutta method maintains accuracy in the BTZ black hole metric. 
Furthermore, we apply the same analysis to the AdS soliton metric, elucidating the difficulties in reconstructing the metric of the AdS soliton using machine learning.


\subsubsection{Exact Solution}
\label{sec:HighTemperatureExact}

In the high temperature phase, the BTZ black hole metric is selected when the bulk is the Einstein theory:
\begin{align}
\label{eq:BTZ_metric}
&ds^2_{\rm BTZ} = -\frac{r_h^2\xi}{L^2(1-\xi)}dt^2 + \frac{L^2}{4\xi(1-\xi)^2}d\xi^2 + \frac{r_h^2}{L^2(1-\xi)}d\theta^2,
\end{align}
This metric leads to the exact solution of the EOM \eqref{eq: EOM before NN} satisfying the ingoing boundary condition ($\xi=0$)\footnote{
The ingoing boundary condition corresponds to the retarded propagation.
} on the horizon as
\begin{align}
    \label{eq: ExactEq}
    \Phi_n &= \xi^{\frac{\alpha_n+\beta_n}{2}}F(\alpha_n,\beta_n,\gamma_n;\xi), 
\end{align}
where F is the Hypergeometric function, and $\alpha_n$, $\beta_n$ and $\gamma_n$ are defined as
\begin{align}
    \alpha_n &:= -i \frac{L^2}{2r_h}(\omega + k_n),~ \beta_n := -i \frac{L^2}{2r_h}(\omega - k_n),\nonumber\\
    \gamma_n &:= 1+\alpha_n+\beta_n.
\end{align}
In this expression, $L$ is the AdS radius and $r_h$ is the horizon radius,  related to the system temperature $T$ as
\begin{align}
    r_h = 2\pi L^2 T.
\end{align}
The temperature $T$ is controllable in the experiment.
We set $L = r_h = 1.00$ for the numerical analysis below, and $C = 2r_h^2/L^2$ in \eqref{eq: gauge condition} for the simplicity of the comparison between the true value of $(\time,\Theta)$ and the learned one.


On the other hand, in the low temperature phase, the AdS soliton metric is selected when the bulk is the Einstein theory:
\begin{align}
&ds^2_{\rm Sol} = -\frac{r_s^2}{L^2(1-\xi)}dt^2 + \frac{L^2}{4\xi(1-\xi)^2}d\xi^2 + \frac{r_s^2\xi}{L^2(1-\xi)}d\theta^2.
\end{align}
In this expression, $L$ is the AdS radius and $r_s$ is the gap of the AdS soliton, related to the ring circumference $a$ as
\begin{align}
\label{eq:r_s}
r_s =\frac{2\pi L^2}{a}\,.
\end{align}
This metric has the same form of the solution as \eqref{eq: ExactEq} that is finite at $\xi=0$ with $\alpha_n, \beta_n$ written as
\begin{align}
    \alpha_n &:= \frac{L^2}{2r_s}(|k_n| - \omega),~ \beta_n := \frac{L^2}{2r_s}(|k_n|+\omega).
\end{align}

\subsubsection{Two solvers}
To illustrate the numerical solvers, we first rewrite the EOM \eqref{eq: EOM before NN} in a vector form:
\begin{align}
    {\bf Z}_n'(\xi) &= {\bf F}(\xi, {\bf Z}_n(\xi))\,, \\
    {\bf F}(\xi, {\bf Z}) &= 
    \left(
    \begin{array}{cc}
        0   &   1  \\
        \omega^2 \Xi(\xi) +k_n^2 \Theta(\xi)  &  -1/\xi
    \end{array}
    \right) {\bf Z} \,.\qquad
    \label{eq:F_xiZ}
\end{align}
with ${\bf Z}_n=(\Phi_n,\Pi_n)^{\rm T}$. 
The Euler method uses the following recurrence relation to determine the values of $\Phi_n$ and $\Pi_n$:
\begin{align}
    {\bf Z}_n &(\xi+\Delta\xi) = 
    \left(
    \begin{array}{cc}
        1   &  \Delta\xi  \\
         \Delta\xi(\omega^2 \Xi(\xi) +k_n^2 \Theta(\xi))   &  1 -  \Delta\xi /\xi
    \end{array}
    \right) {\bf Z}_n(\xi)\,.
\end{align}
Here $\Delta\xi$ is the discretization unit (the lattice constant) in $\xi$, which in this study is set to $-10^{-2}$.
The initial value of ${\bf Z}_n$ is set to the value of the exact solution at $\xi=0.99$ and we use the recurrence relation repeatedly to obtain ${\bf Z}_n(\xi=0.1)$.

The recurrence relation of the Runge-Kutta method is
\begin{align}
    {\bf Z}_n(\xi+\Delta\xi) &= {\bf Z}_n(\xi) + \frac{\Delta\xi({\bf F}_1+2{\bf F}_2+2{\bf F}_3+{\bf F}_4)}{6}, 
    \label{eq:RK}
\end{align}
with
\begin{align*}
    {\bf F}_1 &= {\bf F}(\xi,{\bf Z}_n(\xi)),\\ \nonumber
    {\bf F}_2 &= {\bf F}\left(\xi+\frac{\Delta\xi}{2},{\bf Z}_n(\xi)+{\bf F}_1\frac{\Delta\xi}{2}\right),\\ \nonumber
    {\bf F}_3 &= {\bf F}\left(\xi+\frac{\Delta\xi}{2},{\bf Z}_n(\xi)+{\bf F}_2\frac{\Delta\xi}{2}\right),\\ \nonumber
    {\bf F}_4 &= {\bf F}(\xi+\Delta\xi,{\bf Z}_n(\xi)+{\bf F}_3\Delta\xi).\\ \nonumber
\end{align*}
This form of the Runge-Kutta method generally offers higher accuracy with a global error proportional to the fourth power of $\Delta\xi$, making it more efficient for achieving precise solutions.
We use the same initial value as the Euler Method and find ${\bf Z}_n$ for $\xi \in [0.1,0.99]$.

\subsubsection{Comparison}
We evaluate the numerical reproduction accuracy of the above two solvers by comparing them to the exact solution \eqref{eq: ExactEq}.
\begin{figure*}[t]
\centering
\includegraphics[scale=0.4]{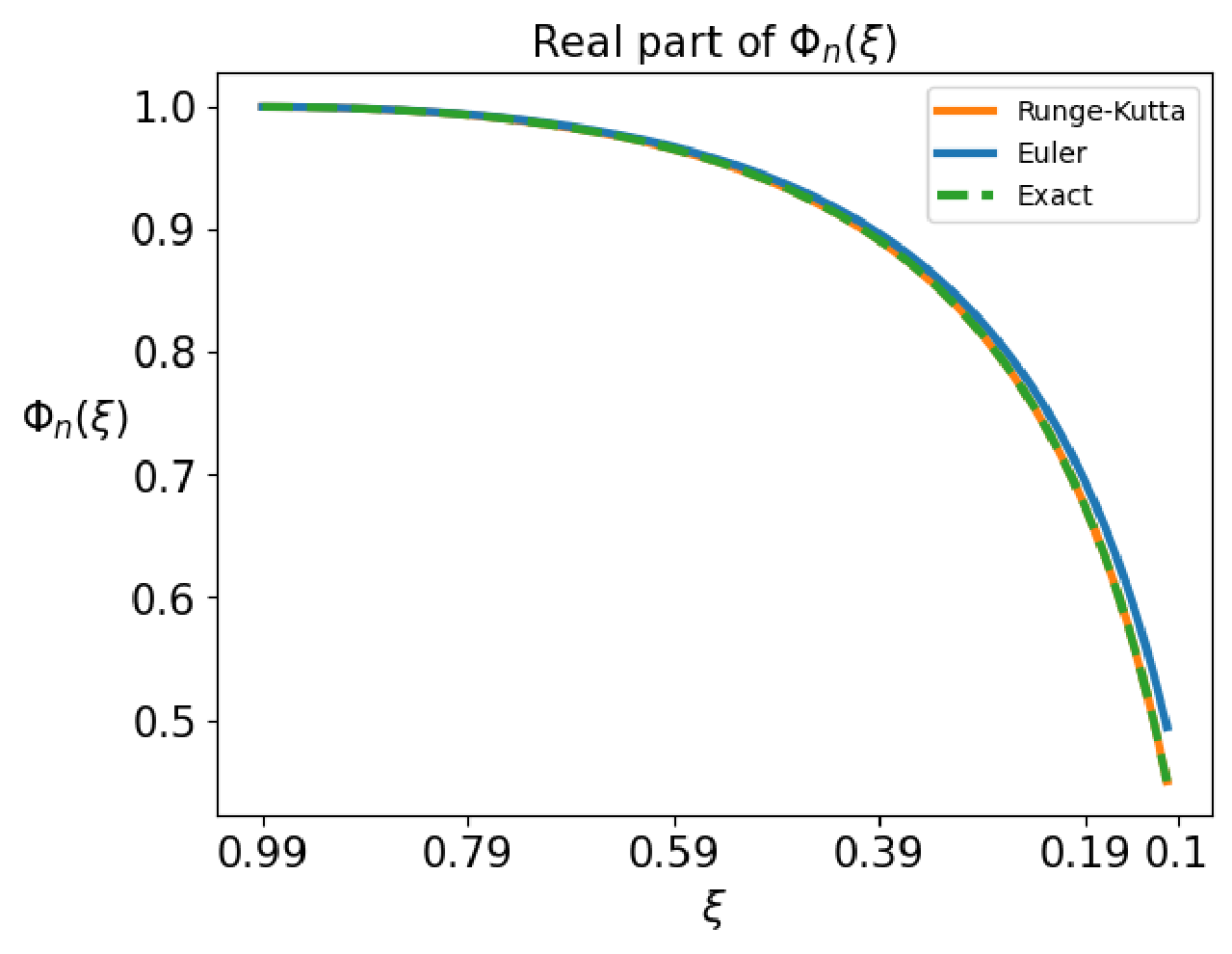}
\caption{The profile of $\Phi_n(\xi)$ on the BTZ black hole metric is shown, comparing the exact solution with numerical results obtained using the Euler method and the Runge-Kutta method. The parameters chosen are $(k_n, \omega) = (1.00, 1.00)$. The blue line represents results from the Euler method, the orange line represents results from the Runge-Kutta method, and the green dotted line represents the exact solution.}
\label{fig:Phi_ncompare}
\end{figure*}
Fig.\ref{fig:Phi_ncompare} presents the numerical calculation results for the real part of $\Phi_n$. As depicted in this figure, the Runge-Kutta method provides a closer approximation to the exact solution compared to the Euler method. Specifically, for the real component of $\Phi_n$ at $\xi=0.1$, the relative error between the Runge-Kutta result and the exact solution is $1.5 \times 10^{-4}$, whereas the relative error between the Euler result and the exact solution is $7.5 \times 10^{-2}$, for the case of $k_n = \omega = 1.00$.
This result shows that the Euler method used in \cite{hashimoto2018deep} is insufficient in the present case for the numerical reconstruction of the differential equation.\footnote{Note that the data structure and the bulk EOM in the present study differ from those in \cite{hashimoto2018deep}: the latter only uses the data at $k_n = \omega = 0$ and employs a nonlinear differential equation. Therefore, making a straightforward comparison with this work using \cite{hashimoto2018deep} as a benchmark is not feasible.} Consequently, as we will describe in the next subsection, we construct a NN model based on the Runge-Kutta method.

Additionally, we perform the same calculations on the AdS soliton metric, which corresponds to the low temperature phase, and compare the two solvers by evaluating their numerical reproduction accuracy.
\begin{figure*}[t]
\centering
\includegraphics[scale=0.4]{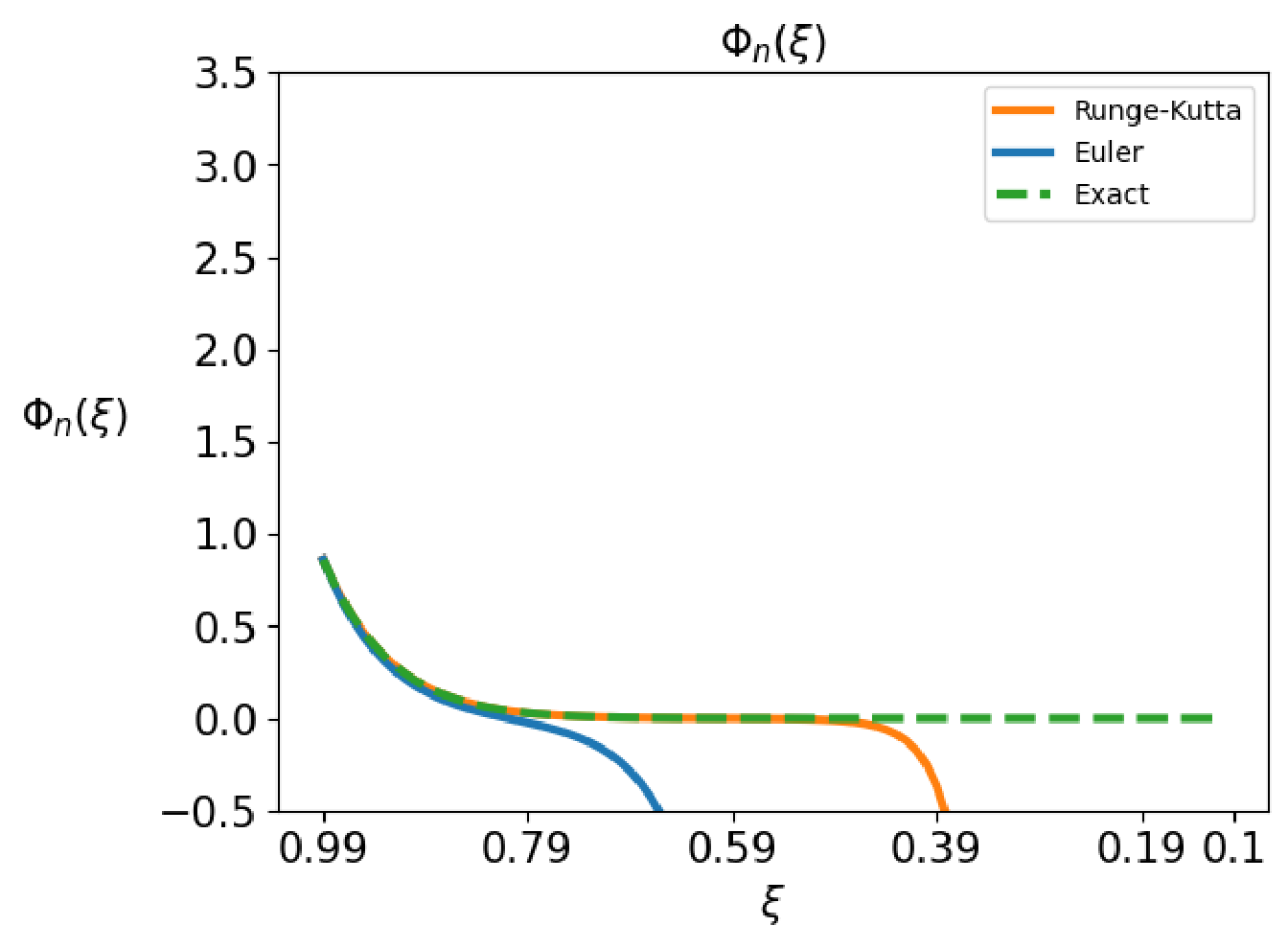}
\caption{The profile of $\Phi_n(\xi)$ on the AdS soliton metric is shown, comparing the exact solution with numerical results obtained using the Euler method and the Runge-Kutta method. The parameters chosen are $(k_n, \omega) = (3.00, 3.00)$ and $r_s = 0.10$. The blue line represents results from the Euler method, the orange line represents results from the Runge-Kutta method, and the green dotted line represents the exact solution.}
\label{fig:Phi_ncompare_soliton}
\end{figure*}
It is important to note that the step size of $k_n$ in the AdS soliton background is $r_s$, whereas in the BTZ black hole background, it is independent of $r_h$. 
To obtain a sufficient amount of data for $k_n$ for effective learning (which we describe later), it is necessary to set $r_s$ smaller than $1$. 
For this reason, we set $r_s=0.1$. 
Fig.\ref{fig:Phi_ncompare_soliton} presents the numerical calculation results for $\Phi_n$ on the AdS soliton metric with $\omega=k_n=3.0$. 
From this figure, it is evident that neither the Runge-Kutta method nor the Euler method approximates the exact solution sufficiently well. 
In fact, the relative error between the Runge-Kutta method results and the exact solution is $-2.5\times 10^{23}$, while the relative error between the Euler method results and the exact solution is $-9.1 \times 10^{23}$.
This result shows that the Runge-Kutta method does not accurately reproduce the exact solution in the low temperature phase.
The main difference in the bulk between the two phases is the boundary condition put at $\xi = 0$.
In the higher temperature phase, the point $\xi=0$ corresponds to the event horizon, and the two independent solutions just have the difference of ingoing or outgoing.
In the low temperature phase, on the other hand, there are two independent solutions, the one finite at $\xi=0$ and the one that diverges there, and we have eliminated the latter.
The fact implies that a tiny numerical error could be enlarged near $\xi=0$, leading to the inaccurate result as in Fig.\ref{fig:Phi_ncompare_soliton}.
We expect that this would be a generic phenomenon that happens in other types of SEMs.
Therefore, we conclude that machine learning in the low temperature phase can in general has a risk to generate a wrong metric, and hence we should focus on the high temperature phase.
Practically speaking, when we apply our method to real materials, we start with reconstructing the bulk metric in the high temperature phase, survey other properties in the same phase, then after that, approach the low temperature phase in a different way.
Hereafter, we focus on the high temperature phase, which is the BTZ spacetime when the bulk is governed by the Einstein theory.

\subsection{Interpretable Neural Network model based on Runge-Kutta Method}
\label{sec:RKNN}
\begin{figure}[ht]
    \vskip 0.2in
    \centering   
    \includegraphics[scale=0.3]{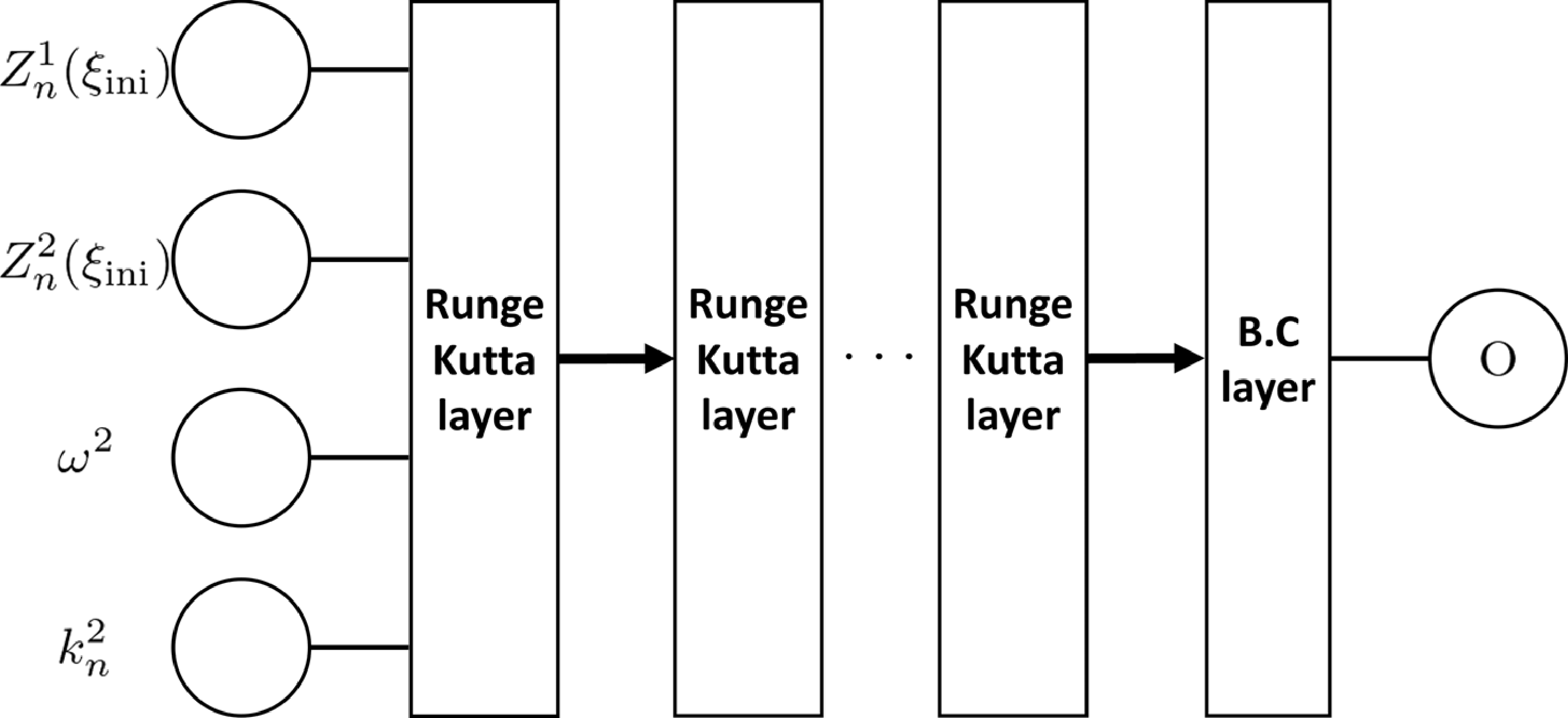}
    \caption{The model structure.}
    \label{fig:NNmodel}
\end{figure}

For the interpretable NN which is identified with the bulk geometry itself, we implement the Runge-Kutta method.
The structure of the NN model is shown in Fig.~\ref{fig:NNmodel}.
In our model, we incorporate the Runge-Kutta layer multiple times, as well as a single boundary condition layer. 
Both of these layers are described in detail in the succeeding subsections.

\subsubsection{\label{R.K Layer}{Runge-Kutta Layer}}
\begin{figure}[ht]
    \vskip 0.2in
    \centering
    \includegraphics[scale=0.3]{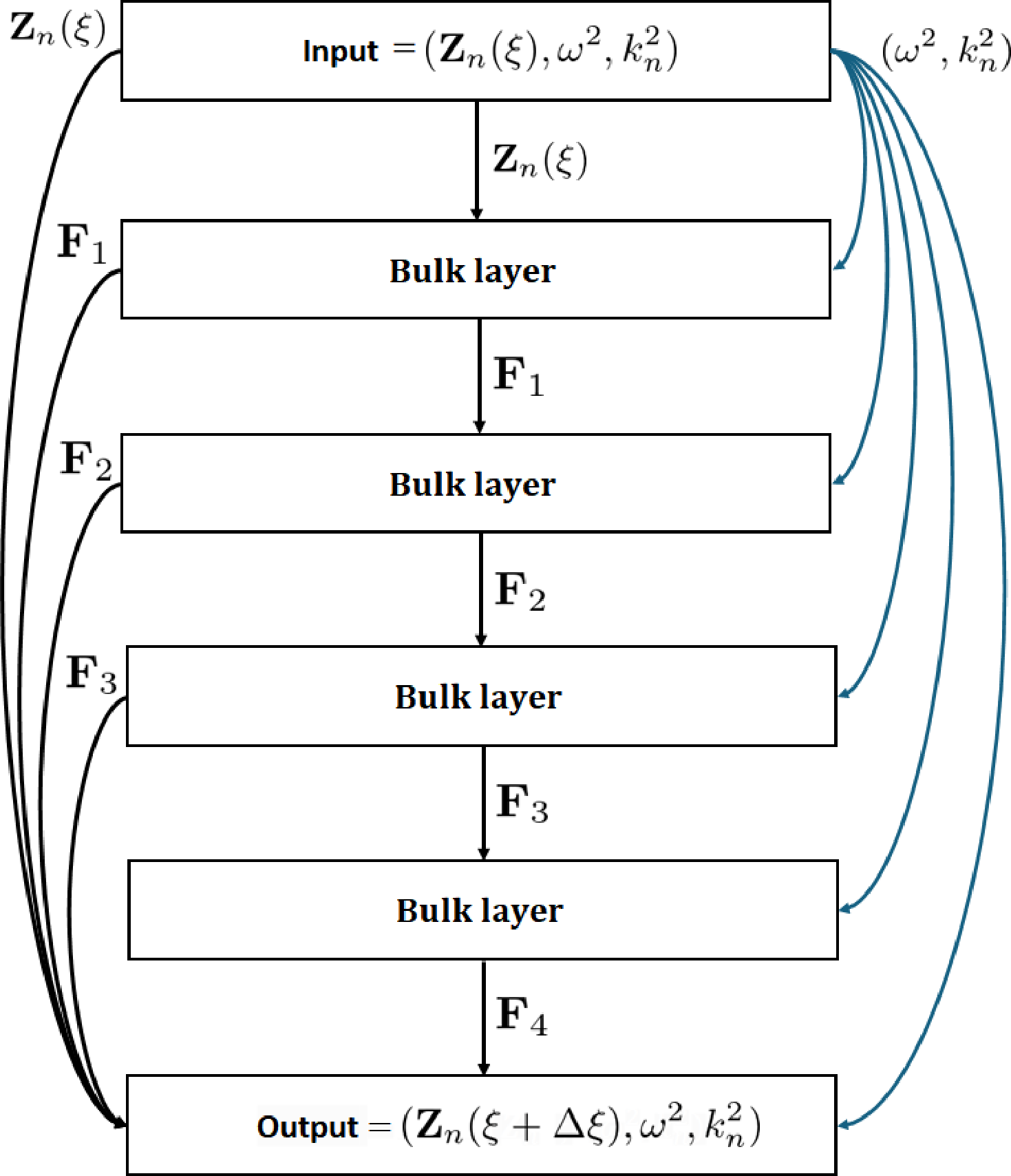}
    \caption{The data flow of the Runge-Kutta layer.}
    \label{fig:RKLayer}
\end{figure}

\begin{figure}[h]
    \vskip 0.2in
    \centering
    \includegraphics[scale=0.2]{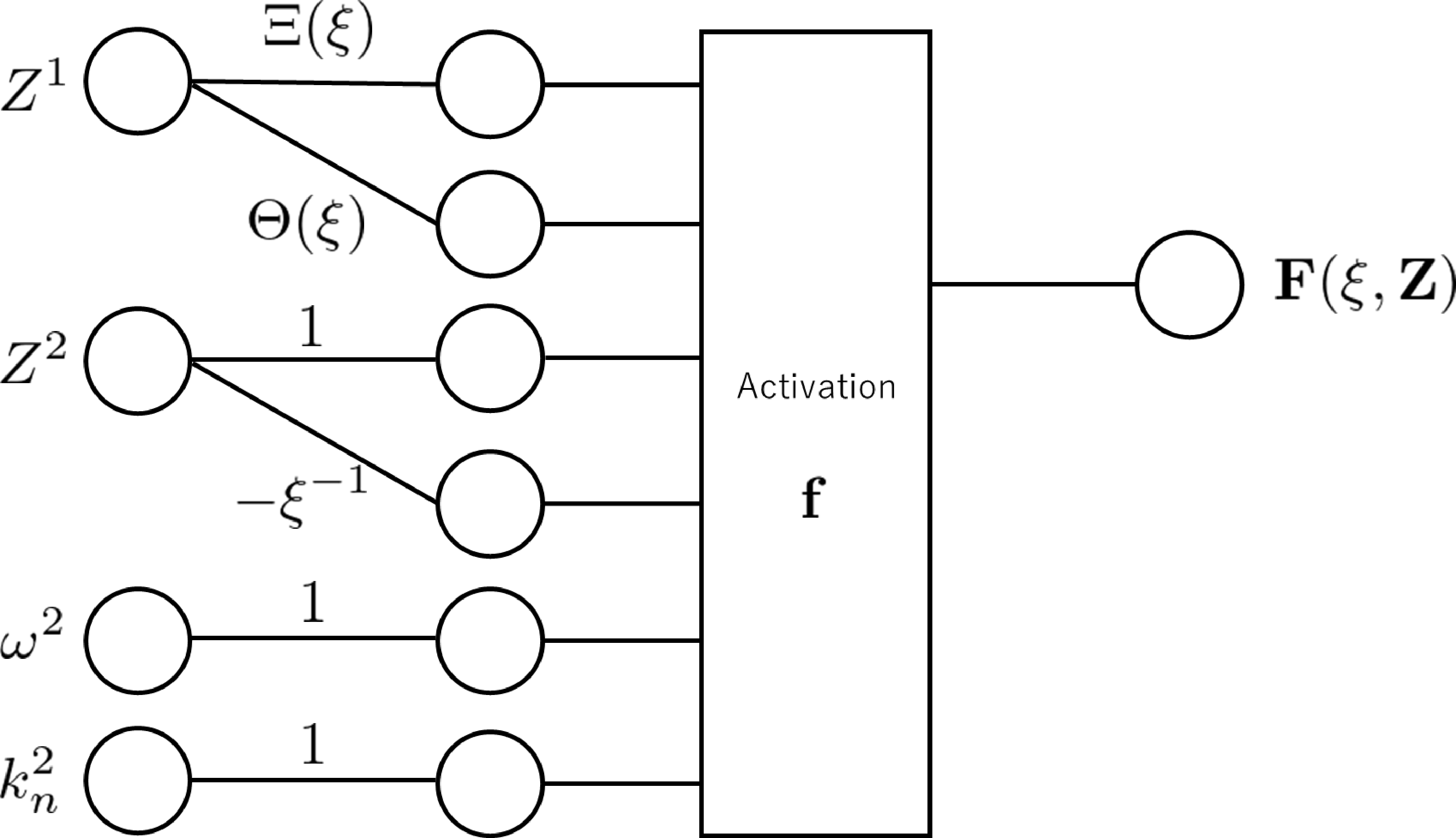}
    \caption{The bulk layer. Here $Z^{1,2}$ is a component of ${\bf Z}$ and the activation ${\bf f}$ is a function of a six dimensional vector ${\bf x}=(x_1,x_2,\ldots,x_6)$ and gives the four dimensional vector $(x_3, x_4+x_5x_1+x_6x_2, x_5,x_6)$.}
    \label{fig:Bulk Layer}
\end{figure}

The implementation of the Runge-Kutta layer is given by \eqref{eq:RK}, where the input and the output of this layer are $({\bf Z}_n(\xi),\omega^2,k_n^2)$ and $({\bf Z}_n(\xi+\Delta\xi),\omega^2,k_n^2)$.
The flow of data in the Runge-Kutta layer is depicted in fig.~\ref{fig:RKLayer}.
The Runge-Kutta layer involves four custom layers, which we call bulk layers, to compute ${\bf F}_1,{\bf F}_2,{\bf F}_3,{\bf F}_4$ in turn.
Fig.~\ref{fig:Bulk Layer} shows the structure of the bulk layer.
The bulk layer receives a four dimensional vector $({\bf Z},\omega^2, k_n^2)$ and returns a two dimensional vector ${\bf F}(\xi, {\bf Z})$ defined in \eqref{eq:F_xiZ}.

In this study, the numerical range of $\xi$ in the radial direction is set from $0.99$ to $0.10$, and the discretization is performed with an incremental width $\Delta\xi$ set to $-10^{-2}$. Therefore, the number of Runge-Kutta layers used in our model is $89$.
In summary, the input of the first Runge-Kutta layer is $({\bf Z}(0.99),\omega^2, k_n^2)$ and the output of the last Runge-Kutta layer is $({\bf Z}(0.1),\omega^2, k_n^2)$, which is the input of the boundary condition layer.

\subsubsection{Boundary Condition (BC) Layer}
As previously mentioned, the behavior of $\Phi_n, \Pi_n$ at the AdS boundary is dictated by the external source and its corresponding response.
In contrast, near the black hole horizon at $\xi = 0$, the behavior is governed by the ingoing boundary condition:
\begin{align}
    \lim_{\xi\to0} \left(
    \xi \Pi_n(\xi) + i\omega \frac{L^2}{2r_h} \Phi_n(\xi)
    \right)
    = 0.
    \label{eq:bc_horizon}
\end{align}
This implies that the response to the external source is configured to satisfy the ingoing boundary condition at $\xi=0$. 
In this study, we aim also to learn the boundary condition in the bulk: to determine the coefficient $i\omega\frac{L^2}{2r_h}$ in \eqref{eq:bc_horizon} though the learning process of NN alongside the bulk metric reconstruction.
Therefore, to reconstruct the boundary conditions through learning, a complex trainable parameter $\rho_n$ is introduced at the deepest layer of the NN to control the boundary conditions:
\begin{align}
    \frac{\xi_{\rm fin}\Pi_n(\xi_{\rm fin})+\rho_n\Phi_n(\xi_{\rm fin})}{\sqrt{|\xi_{\rm fin}\Pi_n(\xi_{\rm fin})|^2+|\rho_n \Phi_n(\xi_{\rm fin})|^2 + \epsilon} } = 0.
    \label{eq:rhon}
\end{align}
The denominator is introduced to compensate the scale of the fields 
and $\epsilon$ is the regularization parameter such that the expression (\ref{eq:rhon}) is numerically well-defined.
In practice, we take $\epsilon = 10^{-6}$.
To be more precise, $\rho_n$ is parameterized as
\begin{align}
    \rho_n = i\omega a_n + |k_n| b_n, 
    \label{eq:rhon_anbn}
\end{align}
with real parameters $a_n$ and $b_n$.
This parameterization incorporates the boundary condition for the BTZ black hole via the following selection:
\footnote{This form of the boundary condition includes that for the AdS soliton by choosing $a_n=0,b_n = - \frac{L^2}{2r_s}$}
\begin{align}
    a_n &= \frac{L^2}{2r_h}\,,\quad b_n = 0 
    \label{eq:BC_BTZ}
\end{align}
In summary, the trainable parameters of our NN model are $\Xi(\xi), \Theta(\xi)$ and a complex parameter $\rho_n$. 

\subsubsection{\label{InitWeight}{Initial Weights}}
\begin{figure*}[t]
\vskip 0.2in
\label{fig:T_AdS}
\includegraphics[scale=0.4]{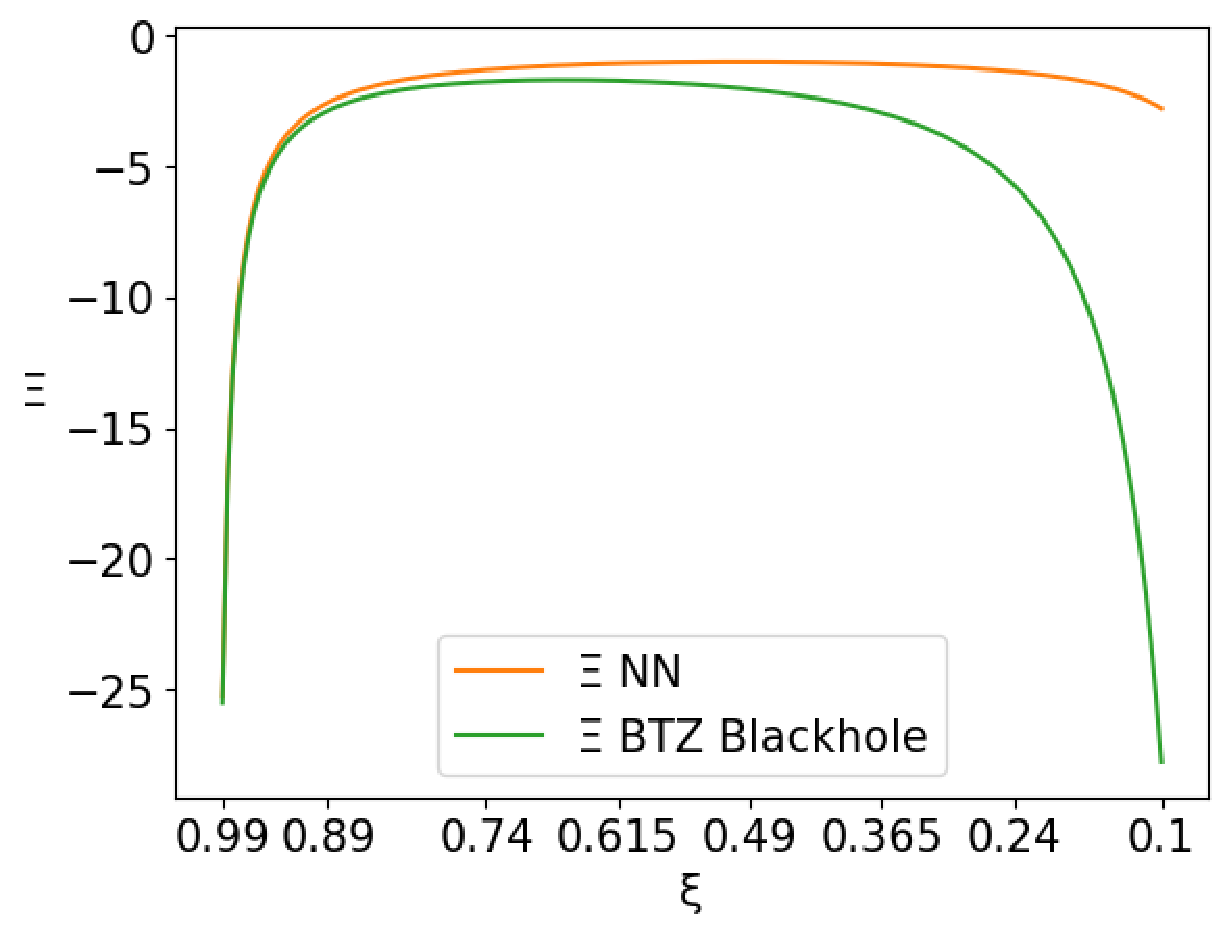}
\includegraphics[scale=0.4]{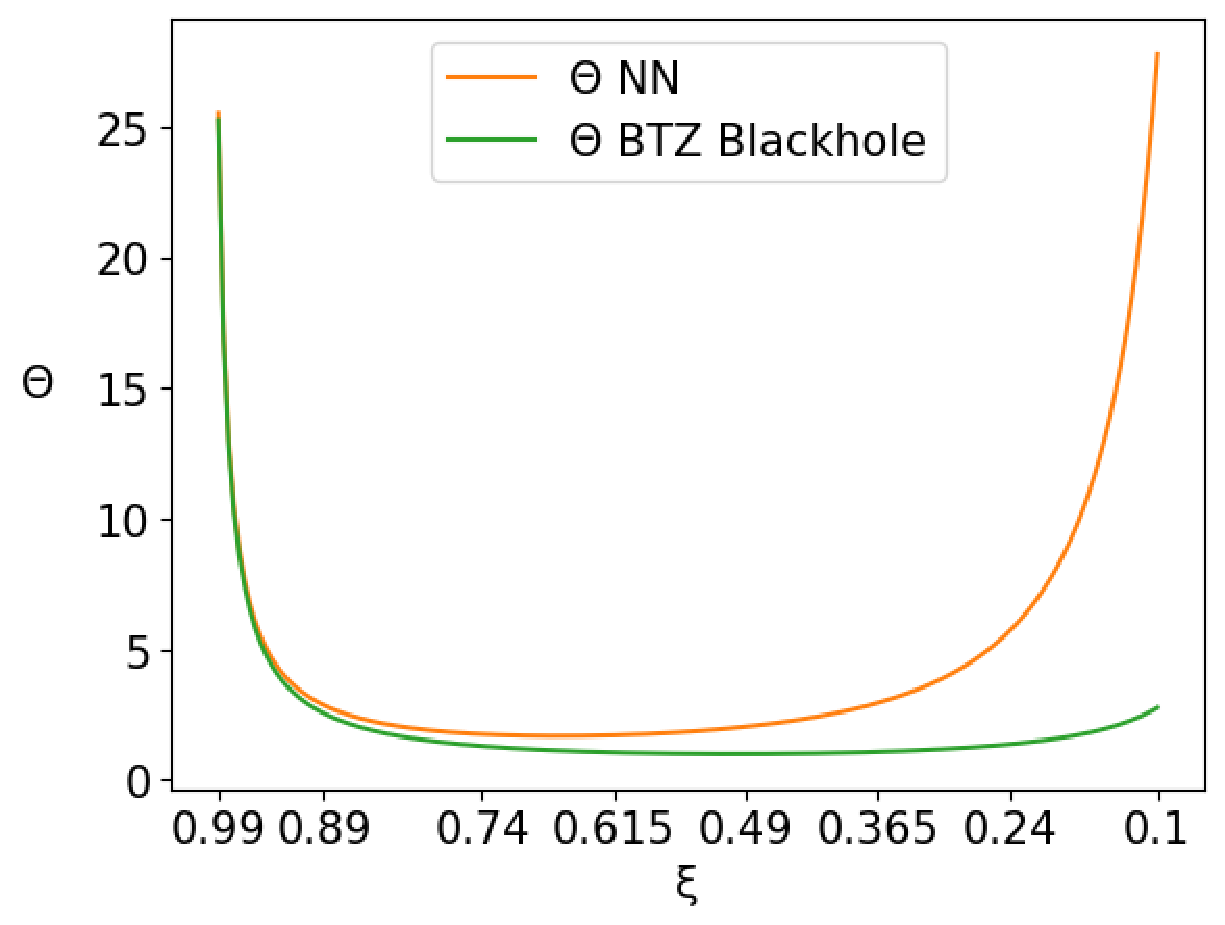}
\caption{\label{fig:THETA_AdS}
Profiles of $(\time,\Theta)$ are shown. The orange lines represent the pure ${\rm AdS}_3$ metric (\ref{eq:AdS_metric}) which is the initial condition for the learning. The green lines represent the ${\rm BTZ}$ black hole metric (\ref{eq:metric}) which is used for generating the training set and also is expected to be reproduced by the learning.}
\end{figure*}

Since our purpose is to reconstruct an asymptotically AdS spacetime\footnote{
``Asymptotically AdS" means that the metric approaches to \eqref{eq:AdS_metric} in $\xi\to 1$.
In AdS/CFT correspondence, this is assumed most of the cases.
}, it is reasonable to choose the initial weights as those of the pure $\mathrm{AdS}_3$ spacetime:
\begin{align}
\label{eq:AdS_metric}
    ds^2_{\rm AdS_3} = -\frac{1}{L^2(1-\xi)}dt^2 + \frac{L^2}{4\xi(1-\xi)^2}d\xi^2 + \frac{\xi}{L^2(1-\xi)}d\theta^2.
\end{align}
This metric leads to $(\Xi,\Theta)$ for pure AdS spacetime as
\begin{align}
    \time(\xi) = - \frac{L^4}{4\xi(1-\xi)}\,,\quad
    \Theta(\xi) = \frac{L^4}{4\xi^2(1-\xi)}
\end{align}
In Fig.~\ref{fig:THETA_AdS}, we show the profiles of $(\time,\Theta)$ for the pure AdS spacetime and the ones for BTZ black hole metric.
Moreover, the initial values of the learning parameters $a_n, b_n$ included in the BC layer are set to $\frac{L^2}{2r_h},- \frac{L^2}{2r_h}$ respectively.

\subsubsection{Loss Function}

This study is a supervised learning, and the data has a label 0 or 1 according to whether it satisfies the boundary conditions at horizon $\xi=0$.
To implement this, we adopt the following loss function $L$ for the training,
\begin{align}\label{eq: Loss function}
L(t) &= -t_{\rm data} \log t - (1-t_{\rm data})\log (1-t), 
\\
t &= \frac12[\tanh\left(100({\rm O}-0.1)\right) - \tanh\left(100({\rm O}+0.1)\right)+2],    
\label{eq:tanaka}
\\
{\rm O} &= \left| \frac{\xi_{\rm fin}\Pi_n(\xi_{\rm fin})+\rho_n\Phi_n(\xi_{\rm fin})}{\sqrt{|\xi_{\rm fin}\Pi_n(\xi_{\rm fin})|^2+|\rho_n \Phi_n(\xi_{\rm fin})|^2 + \epsilon} }\right| \,,
\end{align}
where ${\rm O}$ is the output of the BC layer and is the LHS of \eqref{eq:rhon}. 
Notice that $\Phi_n(\xi_{\rm fin})$ and $\Pi_n(\xi_{\rm fin})$ are obtained from the output of the last Runge-Kutta layer and $\rho_n$ is parameterized as \eqref{eq:rhon_anbn}.
$t_{\rm data}\in \{0,1\}$ is the ground truth label representing whether the ingoing boundary condition is satisfied or not. 
Detailed explanation on the determination of $t_{\rm data}$ is provided in the following subsection.
As a result of \eqref{eq:tanaka}, $t$ is close to 0 if $O$ is sufficiently small but it suddenly approaches $1$ if $O$ moves away from $0$.

\subsection{Dataset}
\label{sec:dataset}

As we shall explain in this subsection, the element of our dataset consists of ${\bf Z}_n(\xi_{\rm ini}), \omega^2, k_n^2, t_{\rm data}$.
Here we describe how to determine the value of $t_{\rm data}$.

Let us emphasize that in this study, we will demonstrate if the learning of our NN model reproduces the BTZ black hole metric in the higher temperature phase.
On the background of BTZ black hole, the exact solution to the EOM (\ref{eq: EOM before NN}) is a linear combination of two independent solutions \cite{hashimoto2022spacetimeemergent}:
\begin{align}
    \Phi_n(\xi) &= C_n^1\xi^{\frac{\alpha_n+\beta_n}{2}}F(\alpha_n,\beta_n,\gamma_n;\xi) \nonumber\\
    &+ C_n^2\xi^{-\frac{\alpha_n+\beta_n}{2}}F(-\beta_n,-1-\alpha_n,1-\alpha_n-\beta_n;\xi).
\end{align}
Here $C_n^{1}$ (or $C_n^2$) is a coefficient of the solution that satisfies ingoing (or out-going) boundary condition.
When we expand this solution around the AdS boundary ($\xi=1$), we have
\begin{align}
    \Phi_n(\xi) &\sim D_n^1(1+\alpha_n\beta_n(1-\xi)\ln(1-\xi) \nonumber \\
    &-(1+\alpha_n\beta_n)(1-\xi))+ D_n^2(1-\xi)/r_h^2, \nonumber \\
    \Pi_n(\xi) &\sim D_n^1(1-\alpha_n\beta_n\ln(1-\xi))-D_n^2/r_h^2,
    \label{eq:phipi_dn}
\end{align}
with 
\begin{align}
    \label{eq:DfromC}
    D_n^1 =& (C_n^2\Gamma(1-\alpha_n-\beta_n)) / (\Gamma(1-\alpha_n)\Gamma(2-\beta_n))) \nonumber\\
    &+ ((C_n^1\Gamma(1+\alpha_n+\beta_n)) / (\Gamma(1+\alpha_n)\Gamma(1+\beta_n)), \nonumber\\
    D_n^2 =& \frac{1}{2}r^2\frac{C_n^2(2+\alpha_n-\beta_n)\Gamma(1-\alpha_n-\beta_n)}{\Gamma(1-\alpha_n)\Gamma(2-\beta_n)} \nonumber\\
    &+ C_n^1\Gamma(1+\alpha_n+\beta_n)\{2-\alpha_n-\beta_n+4\gamma\alpha_n\beta_n \nonumber\\
    &\quad +2\alpha_n\beta_n\psi(1+\alpha_n) 
    +2\alpha_n\beta_n\psi(1+\beta_n) / \Gamma(1+\alpha_n)\Gamma(1+\beta_n)\}. 
\end{align}
Here $\gamma$ is the Euler’s constant and $\psi$ is the digamma function.
According to the AdS/CFT dictionary, $D_n^1$ is regarded as the source of the force exerted on the SEM, and $D_n^2$ is the response of it.

Furthermore, for this to be exactly a solution in all regions of the spacetime, the solution needs to satisfy the boundary condition at the black hole horizon, which is the ingoing boundary condition. So, the outgoing term of this solution should vanish: $C_n^2 = 0$.

\begin{figure*}[t]
    \centering
    \vskip 0.2in
    \hspace*{11mm}
    \includegraphics[scale=0.31]{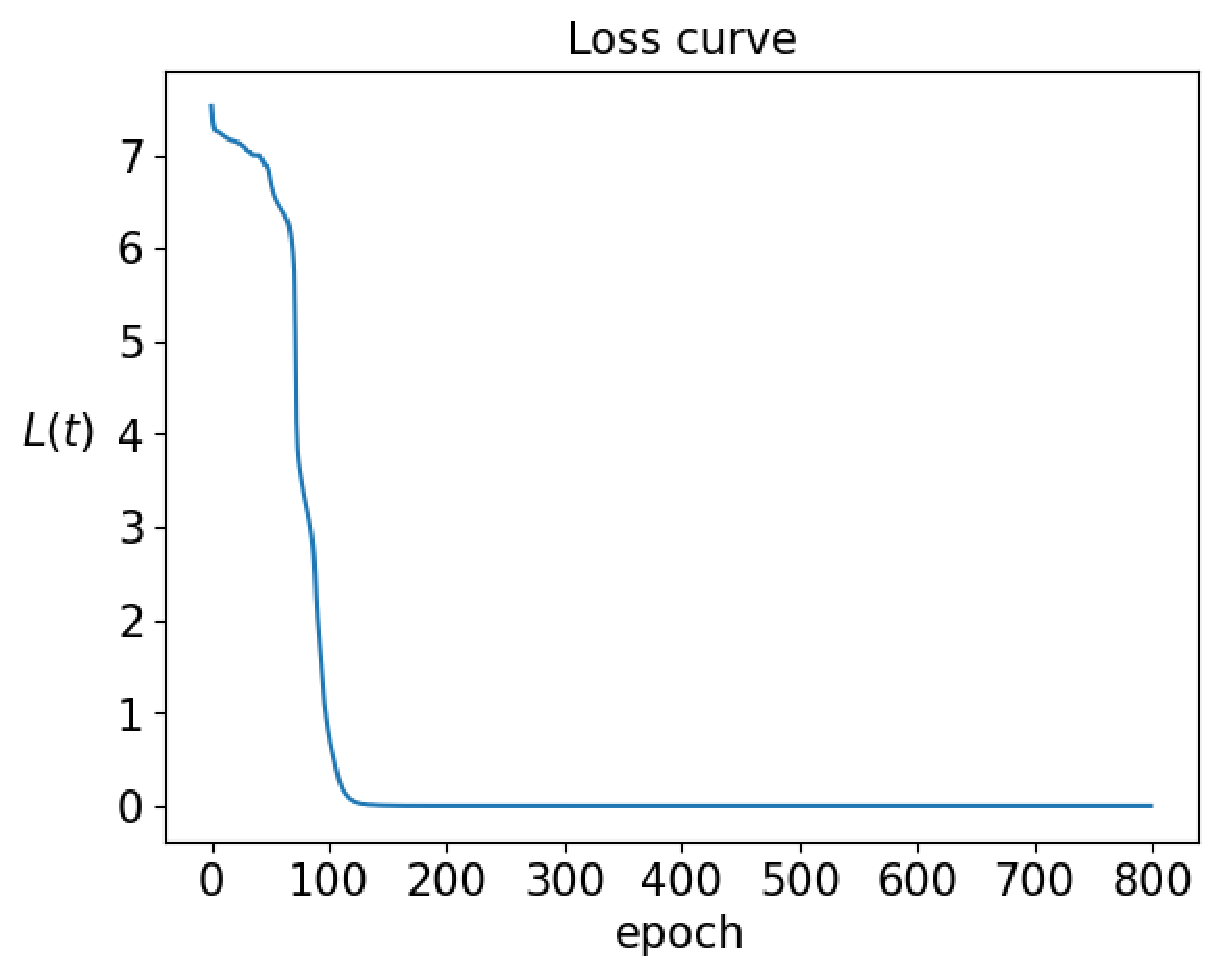}
    \hspace*{0mm}
    \includegraphics[scale=0.31]{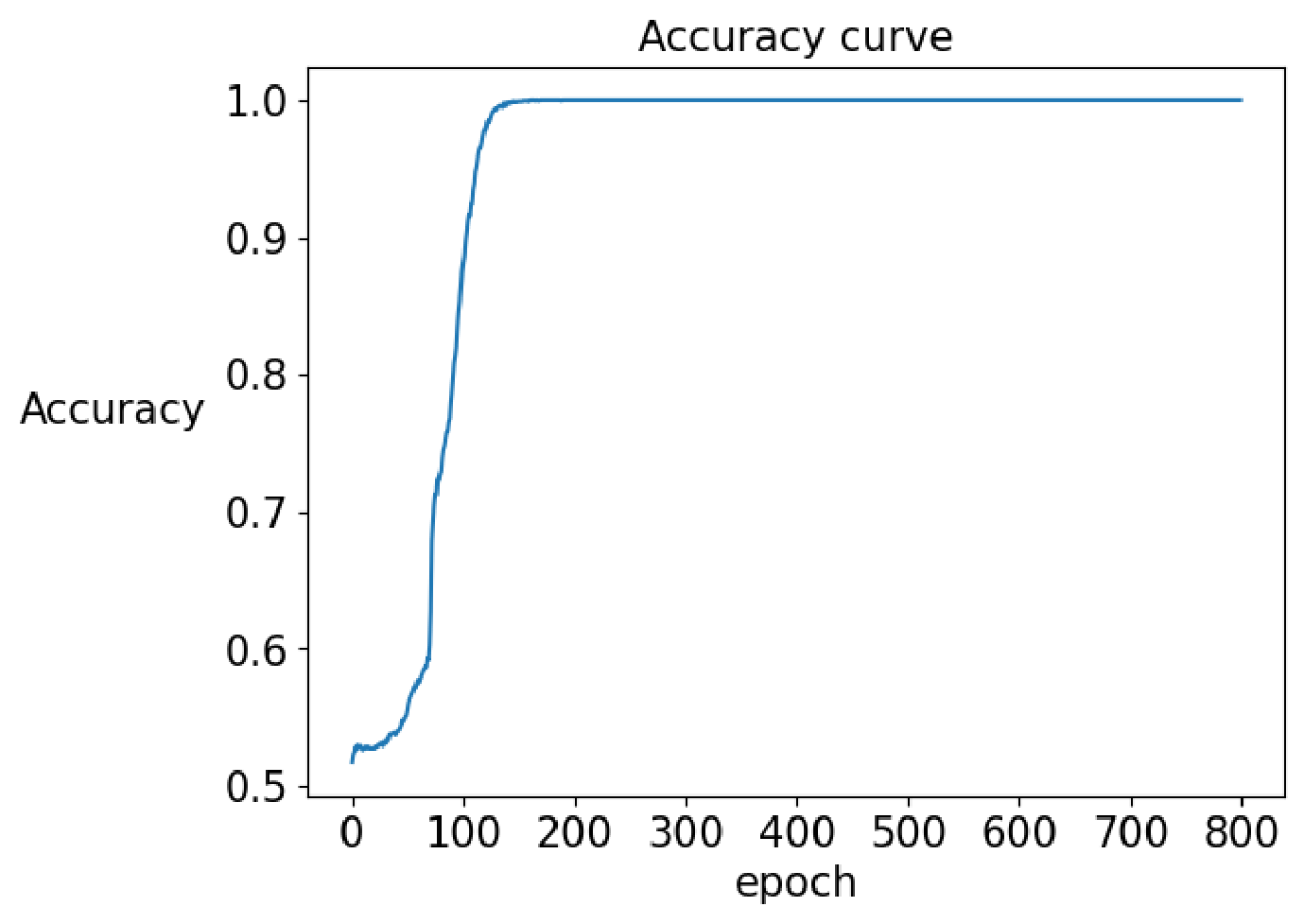}
    \includegraphics[scale=0.31]{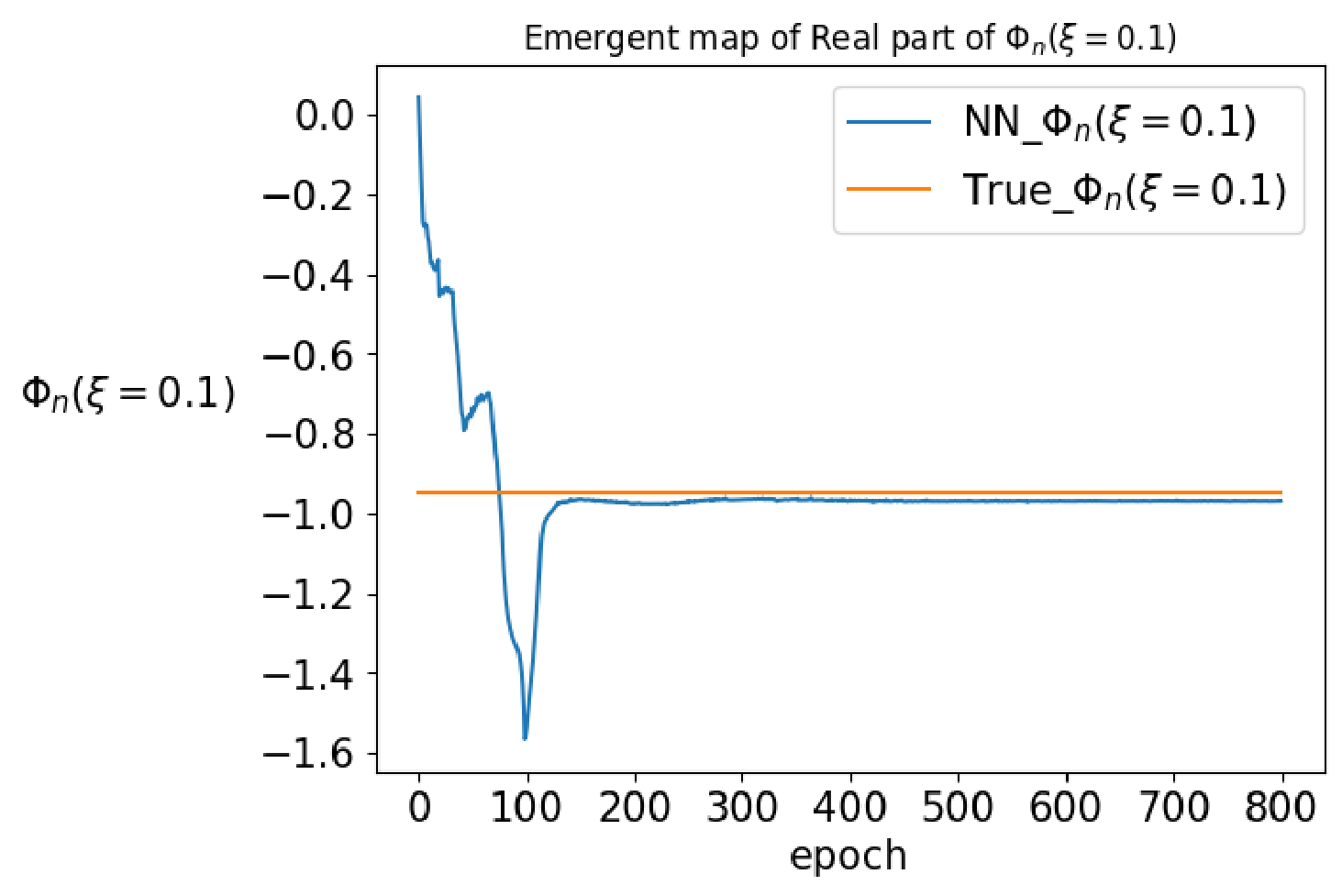}
    \includegraphics[scale=0.31]{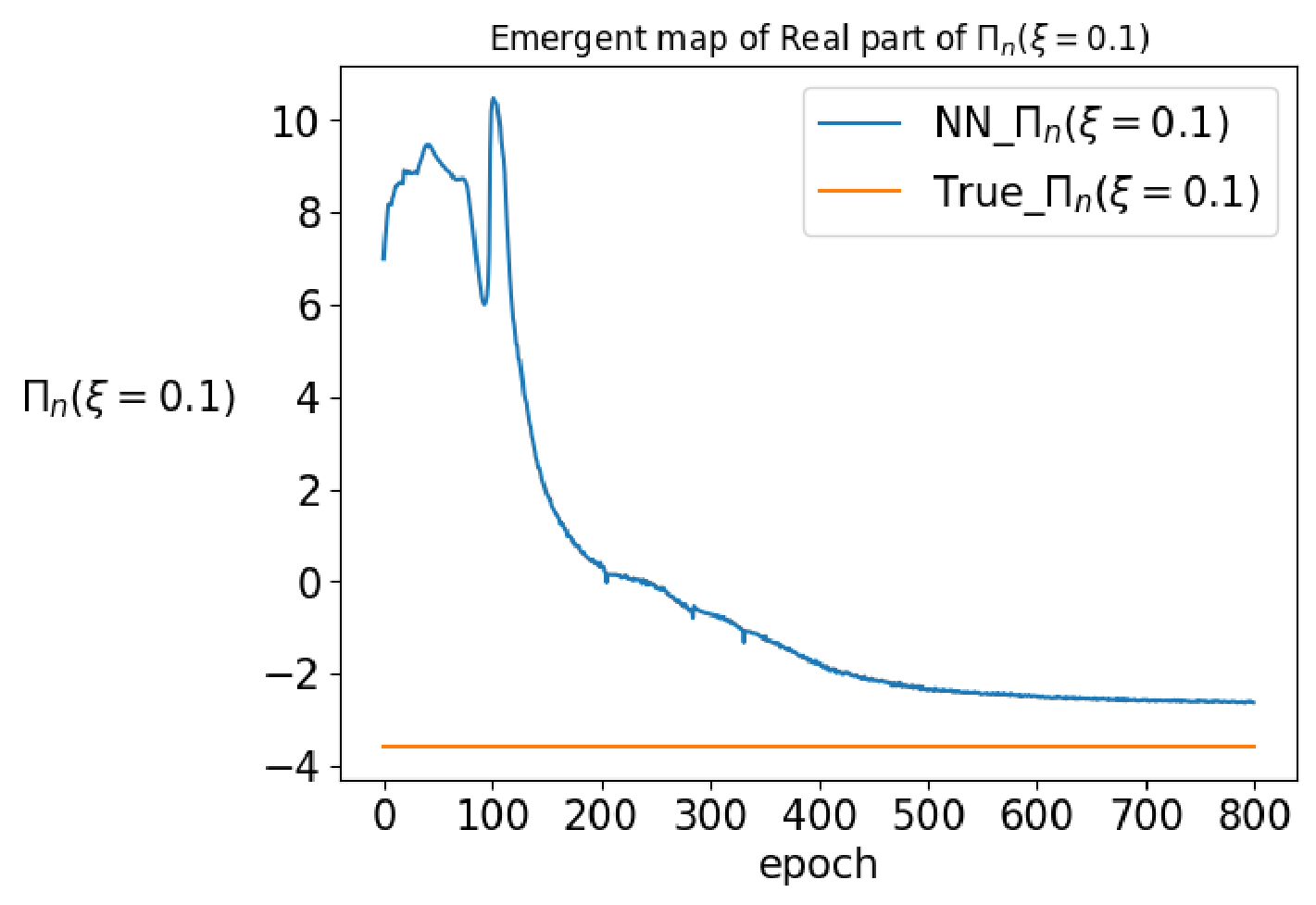}
    \caption{Top: The loss curve and the accuracy curve during the training. Bottom: evolution of the scalar field values during the training. The left (right) panel is the value of the real part of $\Phi$ ($\Pi$) at $\xi=0.1$ for each epoch of the training. The boundary value of the field is chosen to be that of the exact solution for $\omega=3.0,k_n=1.0$. 
    }
    \label{fig:evol}
\end{figure*}

Taking the above into consideration, the following steps are used to generate the dataset.
\begin{enumerate}
    \item Generate $C_n^{1,2}$ randomly. We randomly sample the real and imaginary parts of $C_n^1$ from the uniform distributions ${\cal U} (0.01, 1.0)$ respectively. We adopt $C_n^1$ when $0.5<|C_n^1|<1.2$ otherwise we abandon it. On the one hand, the real and imaginary parts of $C_n^{2}$ are randomly sampled from ${\cal U} (0.001, 0.5)$.
    \item Randomly drown $\omega, k_n$ from the uniform distribution ${\cal U} (0.0, 3.0)$ and ${\cal U} (0.0, \omega)$ respectively.\footnote{
    When $k_n > \omega$, the numerical solution of the scalar field on BTZ metric turns out to have a significant difference from the exact solution. To avoid this, $k_n$ is set to be smaller than $\omega$.
    In addition, we consider the case where the ring circumference $a$ is sufficiently large such that $k_n$ can take any numerical value, not just specific Kaluza-Klein quantized values, as $a$ does not appear anywhere explicitly.}
    \item Classify the positive/negative data by the following decision condition,
    \begin{align}
        t_{\rm data}=\left\{
        \begin{array}{ll}
        0 & (|C_n^2/C_n^1| < 0.01) : {\rm positive}\\
        1 & (|C_n^2/C_n^1| > 1.00) : {\rm negative}
        \end{array}. 
        \right.
    \end{align}
    Here we discard the data if $0.01\leq |C_n^2/C_n^1|\leq 1.0$.
    \item Obtain $D_n^{1,2}$ from $C_n^{1,2}$ using equation (\ref{eq:DfromC}).
    \item Obtain ${\bf Z}_n(\xi_{\rm ini})$ through \eqref{eq:phipi_dn} by substituting $\xi_{\rm ini}$ for $\xi$.
    \item Formulate the data as $\{({\bf Z}_n(\xi_{\rm ini}), \omega^2, k_n^2, t_{\rm data})\}$.
\end{enumerate}
We repeat this procedure until we obtain $1000$ examples for $t_{\rm data}=0$ and $t_{\rm data}=1$ respectively to obtain $2000$ examples in total.

\begin{figure*}[t]
    \vskip 0.2in
    \includegraphics[scale=0.4]{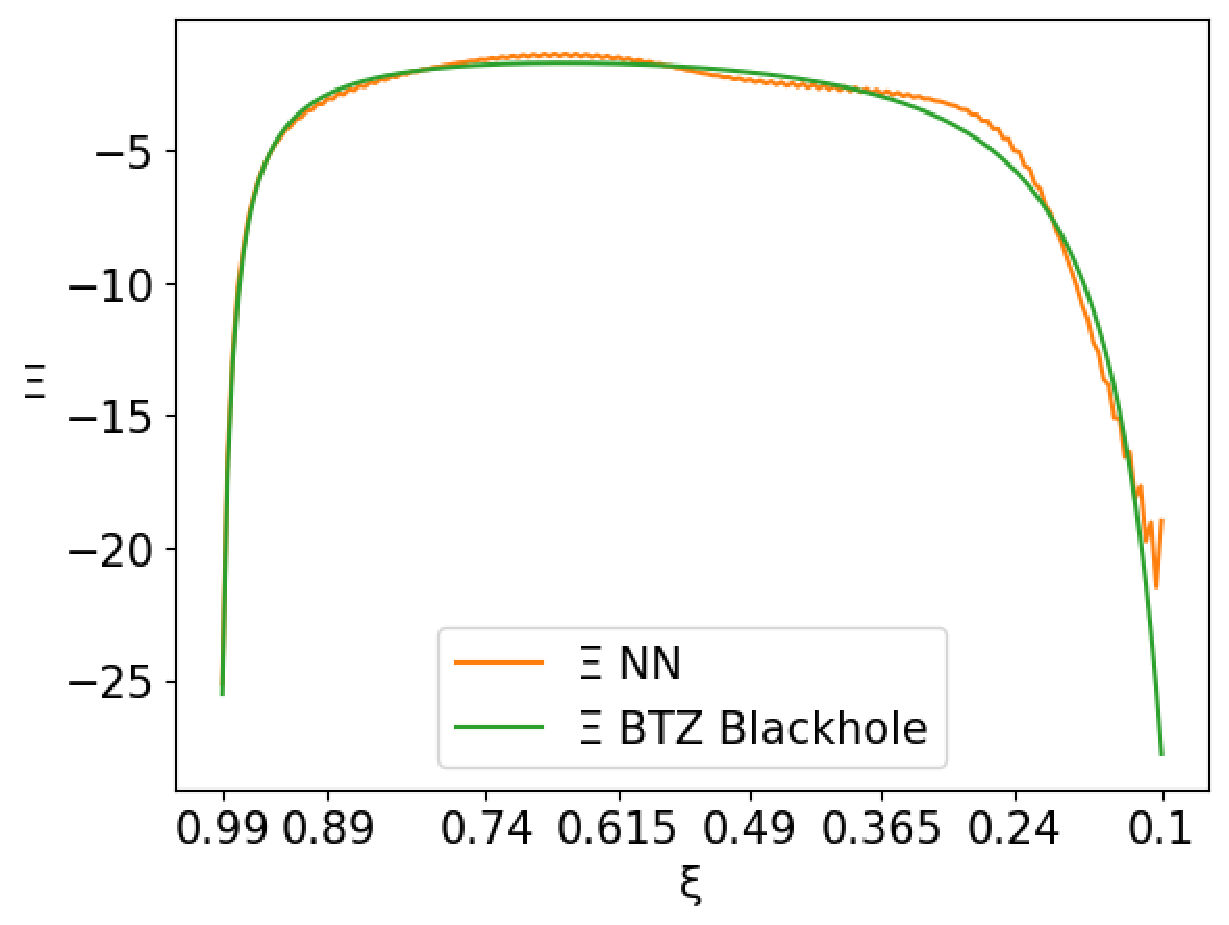}
    \includegraphics[scale=0.4]{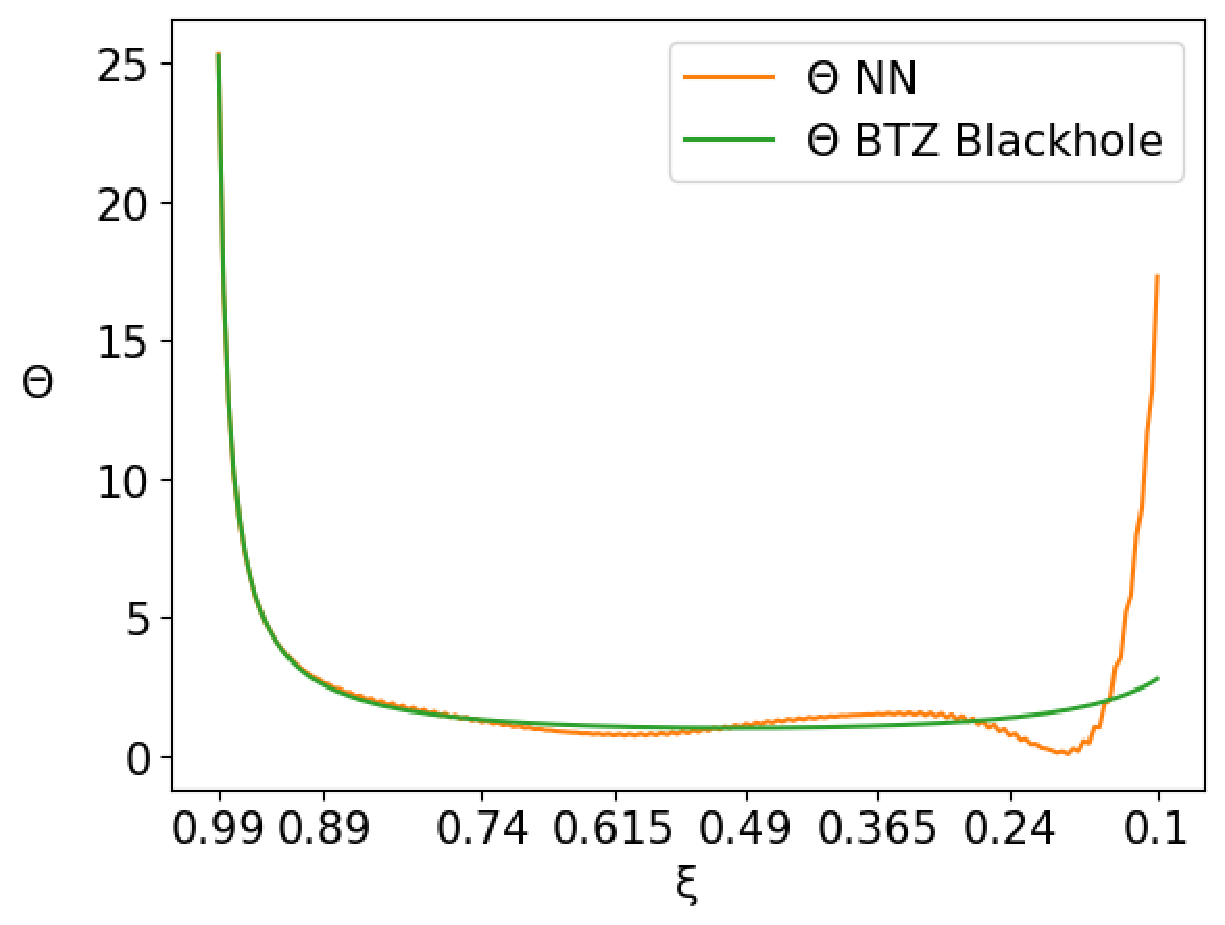}
    \caption{
        Profiles of $\time,\Theta$ after learning. The orange line is the weights of the NN, and the green line is the ground truth, the BTZ black hole metric (\ref{eq:metric}).}
    \label{fig:trained_profiles_noreg}
\end{figure*}

\section{Results of numerical training: emergent spacetime/horizon}
\label{sec:Results}
We build a Neural Network based on the Runge-Kutta method.
The following hyperparameters are commonly used in this study:
optimizer = Adam, batch size = $10$, epoch = $800$. 

Before training, the classification loss value is $7.75$.
After 800 epochs of training, the classification loss value decreases to $-1.02 \times 10^{-7}$, achieving a classification accuracy of $1.0$. 
The learning curves for the loss function and the classification accuracy are depicted in the top of Fig.\ref{fig:evol}.
Additionally, we present the learning curves for the field values at $\xi=0.1$ in the bottom of Fig.\ref{fig:evol}, which shows that the field values closely approach those of the exact solution.

In addition, in this study, the parameter $(a_n, b_n)$ in the expression for the boundary condition is introduced as a learning parameter.
After the learning, this parameter $(a_n, b_n)$ is found to take a value 
$(0.50, 0.01)$. 
If the spacetime is a BTZ black hole, the ingoing boundary condition \eqref{eq:BC_BTZ} results in the values of $(a_n, b_n)$ being $(0.50, 0.00)$. This indicates that we have correctly obtained the ingoing boundary condition at the black hole horizon: {\it The emergence of the black hole horizon is successful.}

Fig.~\ref{fig:trained_profiles_noreg} shows profiles of metric $\time,\Theta$ after learning.
Compared to the initial condition shown by the orange lines in Fig.~\ref{fig:THETA_AdS},
the tendency of the correct BTZ metric is finely reproduced, and the metric is smooth almost everywhere so that we can interpret it as a smooth geometry emergent.

A finer look at the near horizon part of the emergent geometry $\xi \sim 0.1$, we find that the profile of $\Theta$ shown in Fig.~\ref{fig:trained_profiles_noreg} appears to be a little bit deviated from the ground truth. However, this is expected, as this $\Theta$ is related to the angular direction of the metric whose deep IR part near the horizon is largely affected by black hole red shift, meaning that high $\omega$ and large $k_n$ data is needed to probe this part of $\Theta$. Since we used only the range $|\omega|, |k_n| \lesssim {\cal O}(3)$, the difficulty in a fine reproduction of $\Theta$ near the horizon was expected.

\section{Summary and Discussion}

In this paper, in view of future tabletop experiments for quantum gravity through the AdS/CFT correspondence, we build an interpretable neural network architecture
for the bulk reconstruction of the gravity metric by the spacetime-dependent linear response data on a material ring. The ring-shaped material is translation invariant and static at a constant temperature, thus the dual gravity should be
spherically symmetric and stationary spacetime with the topology of a disk times time, which allows two independent metric components to be found. Our machine can spot these components with a good precision thanks to the novel neural network architecture named Runge-Kutta layers. Using a certain set of linear response data, our numerical training of the machine was able to find the BTZ black hole metric, and even discover the black hole horizon condition, meaning that the black hole is emergent in the neural network. Our neural network is ready to be used for real experimental data of linear response on a material ring.

As we described in the introduction, the final goal is to perform a tabletop experiment for quantum gravity, with the discovery of "spacetime-emergent material"
which allows a gravitational holographic description \cite{hashimoto2022spacetimeemergent}. We here make comments 
specifically on our machine compared with the previous study of the bulk reconstruction of the metric.
In this paper, we demonstrate
that the single linear response data of the boundary theory 
which is space- and time-dependent can fix
the two independent bulk metric components, $\Xi$ and $\Theta$, through the machine learning. In the previous studies using machine learning, for example 
in \cite{hashimoto2018deep, hashimoto2018deep2, tan2019deep,yan2020deep, hashimoto2021neural, yaraie2021physics,li2023learning, ahn2024deep}, 
nonlinear response data which is spacetime independent was used to fix a single
component of the bulk metric. As mentioned above, two independent components of the metric need to be found, so our present method is suitable for the bulk reconstruction, generalizing the previous studies.

In general, in the scientific application of machine learning, interpretability is indispensable. While one can obtain solutions to a given equation numerically, if these solutions cannot be meaningfully interpreted by humans, they do not contribute to scientific progress. In this work, we represent differential equations with unknown functional coefficients using a novel neural network architecture and train this network using boundary condition data from the solutions. 
By leveraging physical intuition regarding the sparsity of the neural network, we achieved successful training, and the emergent network provides a meaningful spacetime interpretation. Thus, we effectively solve the problem of reconstructing the differential equation using AI by appropriately incorporating physical biases.

This demonstrates that to achieve physical interpretability in AI, one must impose appropriate physical conditions directly onto the AI architecture. Although it is theoretically possible to perform physical regression from the obtained NN weights, the dimensionality of the NN weight space is typically so large that this approach becomes impractical. Neural ODE \cite{chen2018neural} is no exception.

Considering this, sparse neural networks, such as the recently proposed KANs (Kolmogorov-Arnold Networks) \cite{liu2024kan}, could be promising for future studies. Our study employs the widely used feed-forward neural network with imposed sparsity, showcasing the efficacy of leveraging physical knowledge to enforce sparsity.

The study of AI for problems is not restricted to the interpretable architecture we detailed. Transformers \cite{vaswani2017attention} can be combined to perform mathematical regressions \cite{kamienny2022end, charton2021linear} based on the analogy between mathematical expressions and language sentences \cite{lample2019deep}. Indeed, Transformers have recently been applied to particle theory problems \cite{park2023holography, cai2024transforming, hashimoto2024unification}. Notably, the application of Transformers to decode scattering amplitudes in ${\cal N}=4$ Super Yang-Mills theory \cite{cai2024transforming} closely relates to our subject, as this theory is a prominent model case in the AdS/CFT correspondence.
We expect that combining our current formulation with other machine learning techniques will further enhance AI's capability to solve problems.

\section*{Acknowledgements}
K.~H.~and D.~T.~would like to thank Keiju Murata for valuable discussions.
The work of K.~H., K.~M., M.~M. and G.~O.~was supported in part by JSPS KAKENHI Grant Nos.~JP22H01217, JP22H05111, and JP22H05115.
The work of D.~T.~was supported in part by Grant-in-Aid for JSPS Fellows No.~22KJ1944.

 \newpage
 \appendix
 \onecolumn

\section{On the gauge-fixing}\label{app:gauge-fix}
In this section, we outline the feasibility of implementing the gauge condition \eqref{eq: gauge condition}.

Firstly, we determine the general behavior of $\sqrt{-g}g^{\xi\xi}$ as $\xi$ approaches 1. Given that the metric is asymptotically AdS, its asymptotic form can be derived through a coordinate transformation of \eqref{eq:AdS_metric}. Assuming the metric is static and rotationally invariant as in \eqref{eq:metric}, the remaining gauge freedom is constrained to the transformation $\eta = \eta(\xi)$, where $\eta$ is a smooth bijective function mapping $[0,1]$ to $[0,1]$. Utilizing \eqref{eq:AdS_metric}, the asymptotic form of the metric in $\eta$ coordinates is given by
\begin{align}
    d s^2 \sim -\frac{\alpha}{1-\xi(\eta)}dt^2 + \frac{\xi'(\eta)^2}{4\xi(\eta)(1-\xi(\eta))^2}d\eta^2 + \frac{\alpha\, \xi(\eta)}{1-\xi(\eta)}d\theta^2
    \qquad
    (\eta\to 1),
\end{align}
where $\xi(\eta)$ is the inverse function of $\eta(\xi)$. Here, the gauge transformation and a possible additional constant factor $\alpha$ (as seen in \eqref{eq:BTZ_metric}) are considered. Therefore, in this coordinate system, we have
\begin{align}
   (\sqrt{-g}g^{\eta\eta})^{-1} \sim \frac{\alpha}{2}\xi'(\eta)
   \qquad
    (\eta\to 1).
\end{align}
The right-hand side of this expression may diverge depending on the gauge choice, but the integral remains finite as $\eta \to 1$ due to $\xi(1) = 1$.

Consequently, we can conclude that
\begin{align}
    \int^1 d\xi\,(\sqrt{-g}g^{\xi\xi})^{-1}
    =
    \int^1 d\xi\,\sqrt{\frac{g(\xi)}{f(\xi)h(\xi)}}
    = \text{finite}
    \label{eq: finiteness}
\end{align}
for the general metric described by \eqref{eq:metric}, provided the lower limit is not a singularity. This conclusion is crucial in ensuring the feasibility of the gauge condition.

Next, we examine the limit as $\xi$ approaches 0. The metric behavior in this limit varies depending on the presence of a black hole. In the absence of a black hole, typically the $\mathbb{S}^1$ defined by $t = \text{const}$ and $\xi = \text{const}$ in \eqref{eq:metric} shrinks to zero as $\xi \to 0$, with $f(0)$ remaining finite and non-zero. Thus, we can express $h(\xi) \sim A^2 \xi^{2n}$ for some positive constants $A$ and $n$.

To avoid a conical singularity, the $(\xi,\theta)$-plane must become asymptotically flat in the limit $\xi \to 0$. Using the coordinate transformation $r = \xi^n$, we obtain
\begin{align}
    g(\xi)d\xi^2 + h(\xi)d\theta^2
    \sim
    \frac{g(r^{1/n}) r^{2/n-2}}{n^2}dr^2 + A^2 r^2 d\theta^2
    \qquad
    (\xi \to 0),
\end{align}
where $a$ represents the periodicity of $\theta$. For this expression to be flat, $g(\xi)$ must behave as
\begin{align}
    g(\xi) \sim \left(\frac{A a n}{2\pi}\right)^2 \xi^{2(n-1)}
    \qquad
    (\xi \to 0). \label{eq: g(xi) at origin}
\end{align}
Consequently, there exists a positive constant $B$ such that
\begin{align}
    \sqrt{-g}g^{\xi\xi} = \sqrt{\frac{f(\xi)h(\xi)}{g(\xi)}}
    \sim B \xi
    \qquad
    (\xi \to 0). \label{eq: Y at origin}
\end{align}

If a black hole is present, the same result can be derived with the roles of $f$ and $h$ interchanged. According to our definition of $\xi$, $\xi = 0$ corresponds to the horizon, where $f$ vanishes and is expressed as $f \sim A^2 \xi^{2n}$. The periodicity of $t$ in the Euclidean version of \eqref{eq:metric} corresponds to $T^{-1}$, leading to \eqref{eq: g(xi) at origin} with $a$ replaced by $T^{-1}$. Hence, \eqref{eq: Y at origin} remains valid in this situation as well.

Our objective now is to find a smooth bijective map $\phi: [0,1]\to [0,1]$, $\phi = \phi(\xi)$, such that $\phi(0) = 0$ and $\phi(1) = 1$, satisfying the following condition in $\phi$ coordinates:
\begin{align}
    \sqrt{-g}g^{\phi\phi} = \sqrt{\frac{f(\xi(\phi))h(\xi(\phi))}{g(\xi(\phi))}}\frac{1}{\xi'(\phi)} = C\phi. \label{eq: diff eq}
\end{align}
Here, $C$ must be positive since the left-hand side is positive, and $f$, $g$, and $h$ are the functions described in \eqref{eq:metric}, which represent a generic asymptotically AdS metric satisfying \eqref{eq: finiteness} and \eqref{eq: Y at origin}. Interpreting \eqref{eq: diff eq} as a differential equation, we obtain
\begin{align}
    \phi(\xi) = \exp\left[C \int_1^\xi d\xi'\, \sqrt{\frac{g(\xi')}{f(\xi')h(\xi')}}\right]. \label{eq: phi expression}
\end{align}
By virtue of \eqref{eq: finiteness}, the integrand on the right-hand side is well-defined. In the limit $\xi = \epsilon \ll 1$, \eqref{eq: Y at origin} yields
\begin{align}
    \phi(\epsilon) \sim \exp\left[\frac{C}{B}\int^\epsilon d\xi\,\frac{d\xi'}{\xi'} \right] \sim \epsilon^{C/B},
\end{align}
which is consistent with $\phi(0) = 0$. Furthermore, the right-hand side of \eqref{eq: phi expression} falls within $[0,1]$ and increases monotonically for $\xi \in [0,1]$, as the integrand is uniformly positive. Consequently, \eqref{eq: phi expression} effectively provides a smooth bijection $\phi$ satisfying $\phi(0) = 0$ and $\phi(1)=1$, achieving the transformation to fulfill \eqref{eq: diff eq}.

Under the gauge condition \eqref{eq: gauge condition}, all metric components can be reconstructed from $\time$ and $\Theta$ as
\begin{align}
    f(\xi) = -g_{tt}(\xi) = C\xi^2 \Theta, \qquad
    g(\xi) = g_{\xi\xi}(\xi) = -C\xi^2 \time \Theta, \qquad
    h(\xi) = g_{\theta\theta}(\xi) = -C\xi^2 \time.
\end{align}
\section{\label{sec:Another}Initial weight dependence}
\begin{figure}[h]
    \centering
    \vskip 0.2in
    \includegraphics[scale=0.35]{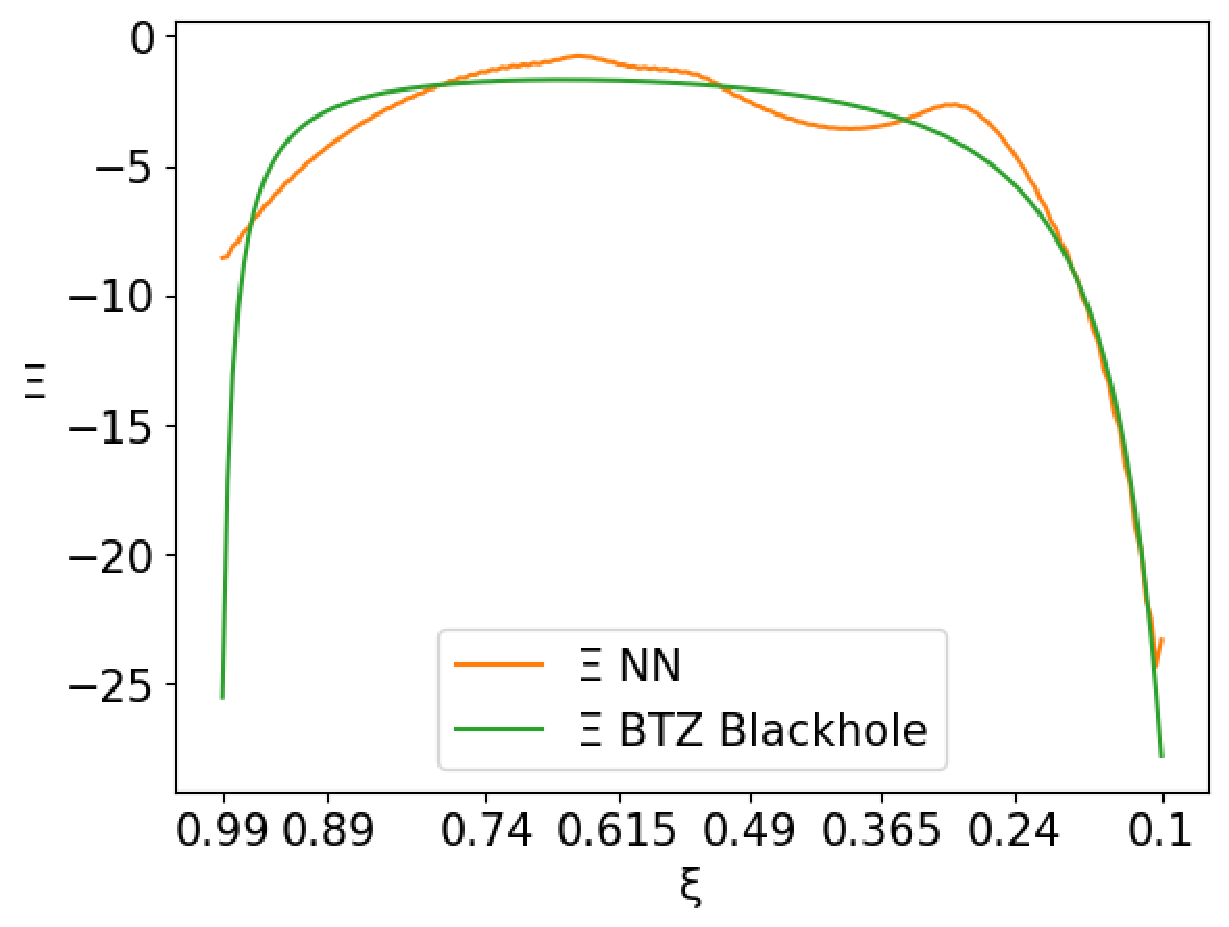}
    \includegraphics[scale=0.35]{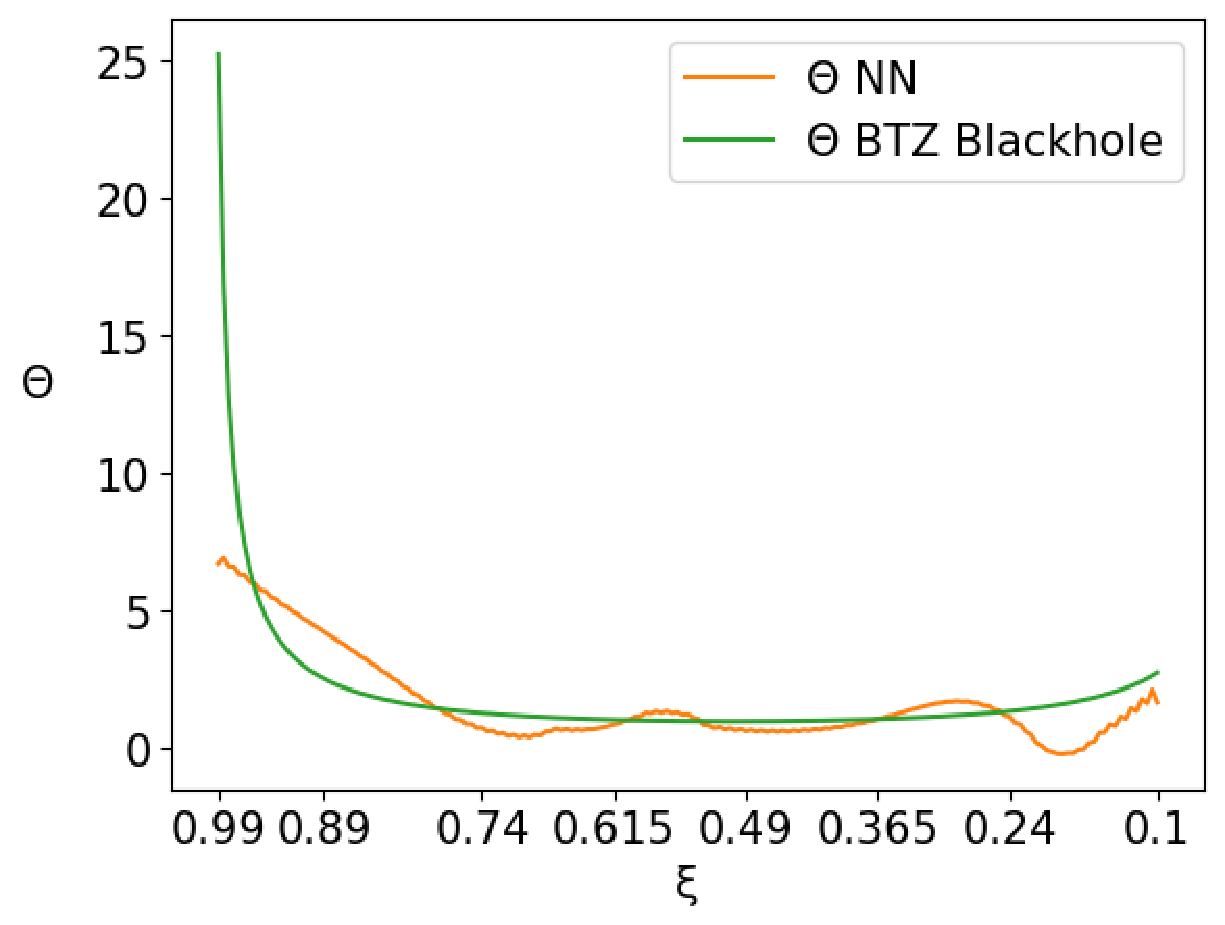}
    \caption{Profiles of $\time$ and $\Theta$ after training with the constant initial setup. The orange line represents the neural network weights, and the green line corresponds to the ${\rm BTZ}$ metric (\ref{eq:BTZ_metric}).}
    \label{fig:trained_profiles_const}
    \vskip -0.2in
\end{figure}

As previously mentioned, the initial setup for the geometry fed into the neural network is the pure AdS spacetime
\eqref{eq:AdS_metric}. In this appendix, we present the training outcomes using an alternative initial setup to verify whether the results we obtained are universal.
For this alternative initial setup, we simply use a constant function for $\time(\xi)$ and $\Theta(\xi)$. 
Employing the same dataset, architecture, and weight update strategies, we obtained the trained weights which are depicted in Fig.~\ref{fig:trained_profiles_const}.

As illustrated in Fig.~\ref{fig:trained_profiles_const}, the resulting metric aligns with the BTZ black hole metric, though the deviation from the ground truth is more significant in comparison to the pure AdS initial setup shown in Fig.~\ref{fig:trained_profiles_noreg}. 
The reason behind this is straightforward: since the BTZ black hole metric is asymptotically AdS, meaning it matches the pure AdS at large $\xi$, training is more straightforward with the pure AdS initial setup.
Nevertheless, it is encouraging to observe that, even with the constant initial setup, the trained metric is confirmed to be consistent with the BTZ black hole metric.

\section{\label{sec:PIReg}Physics-informed regularization}

In our future research, we intend to use actual experimental data from a ring-shaped material, in which case the emergent bulk metric will not necessarily satisfy an Einstein equation. Moreover, for a single data set, there could be multiple emergent spacetimes, contingent on the quantity of data.\footnote{For instance, in our scenario, we have utilized a limited range of $\omega$ and $k_n$. If one employs data with all possible values of $\omega$ and $k_n$, then a singular emergent geometry can be pinpointed. However, with a restrained amount of data, a range of emergent geometries might be learned.}
To isolate a plausible geometry that is physically more understandable, we introduce the concept of free energy regularization to measure the extent to which the emergent geometry adheres to the Einstein equation.
As discussed around \eqref{eq:metric}, the system is assumed to be in thermal equilibrium, which implies that the free energy reaches its minimum at a fixed temperature $T$.

Within the AdS/CFT framework, the leading term of free energy in the string scale expansion in the bulk is known to be represented by the following Euclidean action\footnote{This type of ``Einstein regularization" was described in \cite{hashimoto2018deep} and \cite{hashimoto2019ads}.}:
\begin{align}
    S_{\rm E} &= S_{\rm EH} + S_{\rm GH} + S_{\rm CT}, \nonumber \\
    S_{\rm EH} &= 2\pi\beta\int^{\xi_{\rm ini}}_{\xi_{\rm fin}}d\xi\sqrt{fgh}\left[\frac{1}{L^2}-{\rm Ricci}\right], \quad 
    S_{\rm GH} = \left.-2\pi\beta\sqrt{\frac{fh}{g}}\left(\frac{f'}{f'}+\frac{h'}{h}\right)\right\rvert_{\xi_{\rm ini}}, \quad \nonumber \\
    S_{\rm CT} &= \left.\frac{4\pi\beta}{L}\sqrt{fh}\right\rvert_{\xi_{\rm ini}}, \nonumber \\
    {\rm Ricci} &= \frac{1}{2f^2g^2h^2}(fg'h(fh)' + f^2gh'^2 + f'^2gh^2 
    -2ff''gh^2 -ff'ghh'-2f^2ghh'' ), 
\end{align}
where $S_{\rm {EH}}$ is the Einstein-Hilbert action, ${\rm S_{GH}}$ represents the Gibbons-Hawking-York boundary term \cite{Chakraborty_2017}, and we have introduced
\begin{align}
    f(\xi) &= -g_{tt}(\xi) = -\frac{g_{\xi\xi}(\xi)}{\time(\xi)}, \qquad 
    g(\xi) = g_{\xi\xi}(\xi)  = -\frac{4r_h^4\xi^2\time(\xi)\Theta(\xi)}{L^6}, \qquad \nonumber \\
    h(\xi) &= g_{\theta\theta}(\xi) = \frac{g_{\xi\xi}(\xi)}{\Theta(\xi)}. 
\end{align}
Given the action is Euclidean, we used the Euclidean signature of \eqref{eq:metric}.
Minimizing this action is tantamount to solving the equations of motion, which are the Einstein equations.

Consequently, to guarantee the system's thermal equilibrium, we can introduce the following regularization term:
\begin{align}{\label{eq:FreeEnergy}}
    R_{\rm E} &= S_{\rm E}.
\end{align}
To complete the free energy regularization, the temperature must be held constant.
In the bulk description, the temperature is defined by the metric value at the horizon:
\begin{align}
    T = \frac{1}{4\pi}\frac{f'(\xi_{\rm fin})}{\sqrt{f(\xi_{\rm fin})g(\xi_{\rm fin})}}.
\end{align}
This result is derived by enforcing the regularity of the Euclidean version of \eqref{eq:metric} at $\xi = 0$, which dictates the inverse temperature $T^{-1}$ or the periodicity of Euclidean time.
Thus, besides \eqref{eq:FreeEnergy}, we also need to introduce
\begin{align}\label{eq: T regularization}
    R_{\rm T} = \left(T - \frac{1}{4\pi}\frac{f'(\xi_{\rm fin})}{\sqrt{f(\xi_{\rm fin})g(\xi_{\rm fin})}}\right)^2\,,
\end{align}
as another regularization term.

The NN is trained with the additional regularization terms $R_{\rm E}$ and $R_{\rm T}$. Our earlier results shown in Sec.~\ref{sec:Results} remain largely unchanged, while the loss value of the new regularization terms is significantly small. This indicates that the emergent metric obtained closely aligns with the BTZ black hole, as we previously stated in Sec.~\ref{sec:Results}.

Here is a technical note on the training procedures.
Initially, the NN is trained using 
the loss function \eqref{eq: Loss function}. 
Then, after $400$ epochs, we add \eqref{eq:FreeEnergy} and \eqref{eq: T regularization}.
This is because the initial condition (pure AdS spacetime) is inherently a stationary point of the action $S_\mathrm{E}$. Thus, our selected initial weight in \eqref{eq:AdS_metric} is likely to remain near the initial value in the weight space, potentially hindering successful training of the NN.

\begin{figure}[h]
    \centering
    \vskip 0.2in
    \includegraphics[scale=0.35]{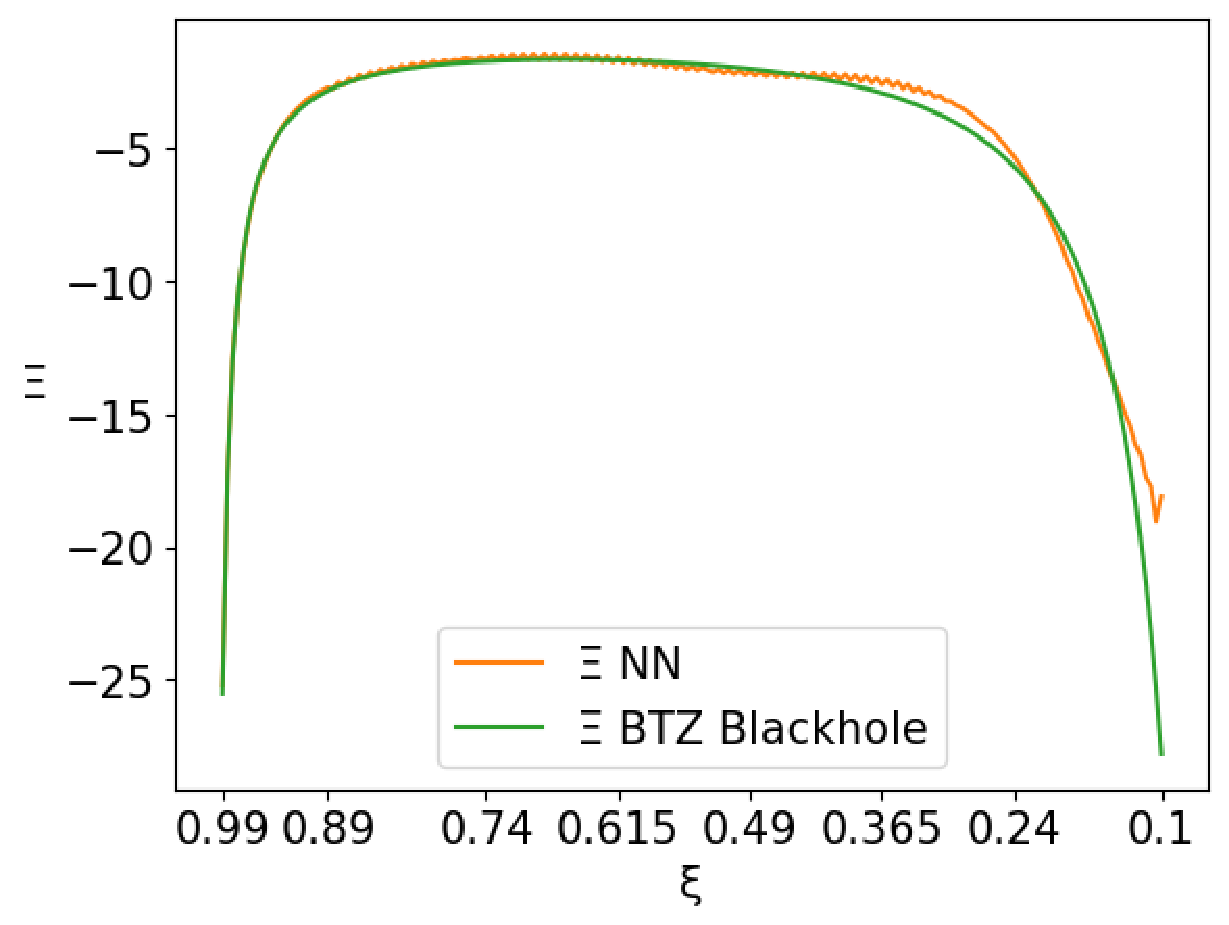}
    \includegraphics[scale=0.35]{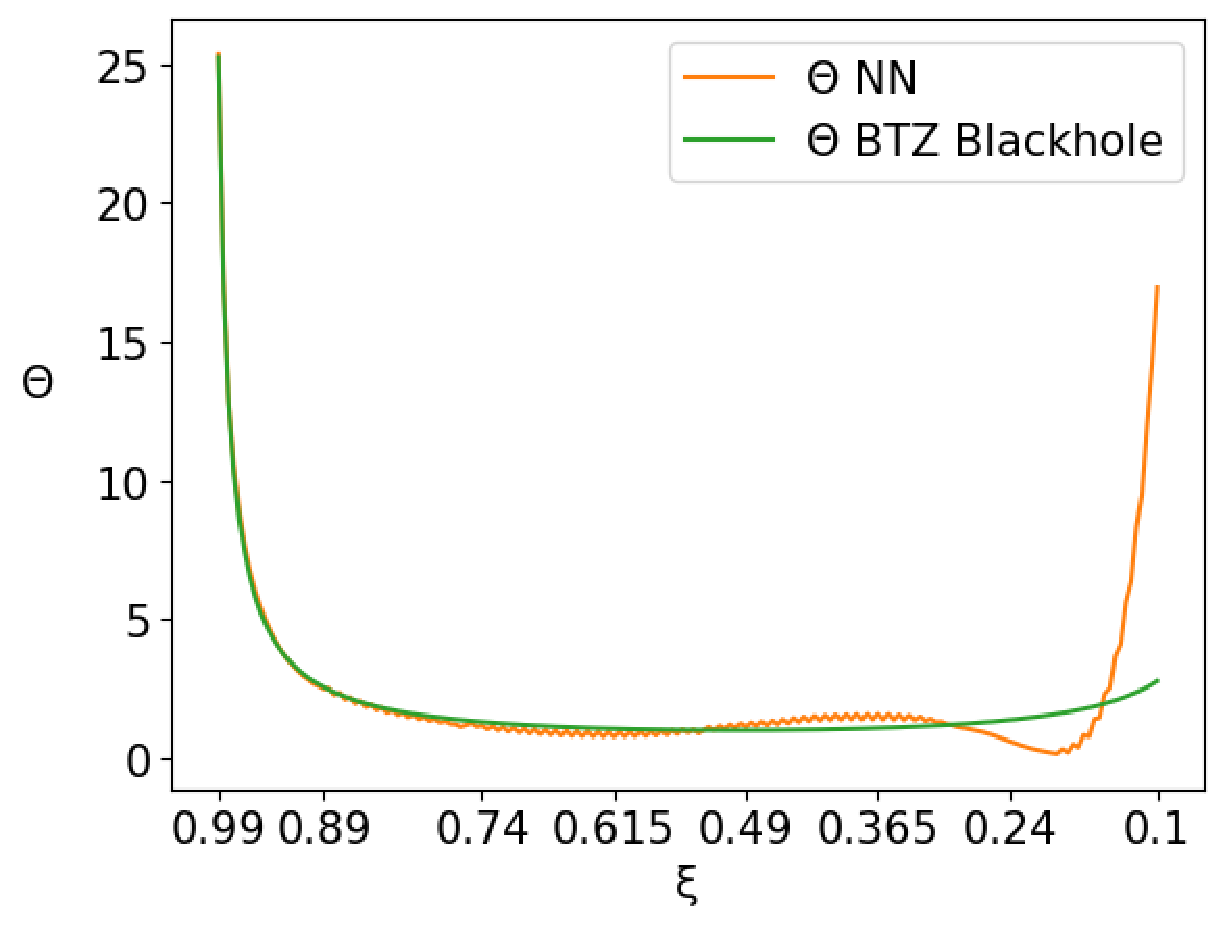}
    \caption{
        Profiles of $\time$ and $\Theta$ after learning with the addition of the physics-informed regularization. The orange line depicts the weights of the NN and the green line shows the ${\rm BTZ}$ metric (\ref{eq:BTZ_metric}).}
    \label{fig:trained_profiles}
    \vskip -0.2in
\end{figure}

After $400$ epochs of training with the added 
regularizations \eqref{eq:FreeEnergy} and \eqref{eq: T regularization}, the total loss stands at 
$1.16$,
the class loss at 
$1.00 \times 10^{-7}$, 
the accuracy at $1.00$, and $(a_n,b_n)$ at 
$(0.50,0.03)$.
These outcomes are on par with those in the previous section, suggesting the emergent geometry is indeed interpretative as a BTZ black hole. Refer to Fig.~\ref{fig:trained_profiles} for the trained profiles of the metric functions.

The assertion above is merely indicative, owing to the following concern.
A potential issue with the temperature regularization term \eqref{eq: T regularization} is its tendency to always be satisfied through adjusting a few near-horizon weights. In fact, detailed comparison of the profiles in Figs.~\ref{fig:trained_profiles_noreg} and  \ref{fig:trained_profiles} reveals that such local tuning likely occurred during training, implying that physically, the regularization \eqref{eq: T regularization} might not have been effective. This issue is planned to be addressed in future work.

\section*{References}
\bibliographystyle{unsrt}
\bibliography{SEM_paper}
\end{document}